\documentstyle[12pt,epsf,cite]{article}
\def\hyp#1#2{${}^{#1}_{\Lambda}$#2}
\def\eqn#1{eq.~(\ref{#1})}
\def\refer#1{ref.~\cite{#1}}
\def\lnnn{$\Lambda N \to N N$ }

\def\L{$\Lambda$\ }

\begin{document}




\title{\bf Weak Decay of $\Lambda$ Hypernuclei}

\vspace{1cm}

\author{E. OSET$^{1)}$\thanks{e-mail:
oset@evalvx.ific.uv.es} and
A. RAMOS$^{2)}$\thanks{e-mail: ramos@ecm.ub.es} \\
{~} \\
{\small \it
\noindent  1) Departamento de F\'{\i}sica Te\'orica and
IFIC}\\
{\small\it \noindent Centro Mixto Universidad de Valencia-CSIC}\\
{\small\it \noindent 46100 Burjassot (Valencia), Spain} \\
{~} \\
{\small\it \noindent 2) Departament d'Estructura i Constituents
de la
Mat\`eria}\\
{\small\it \noindent Facultat de F\'{\i}sica, Universitat de
Barcelona}\\
{\small\it \noindent 08028-Barcelona, Spain}
}

\date{}

\maketitle

\vskip 3cm

\begin{abstract}

We review recent developments concerning the weak decay of
$\Lambda$ hypernuclei. New
studies covering the mesonic decay channel 
as well as recent models for the non-mesonic one are discussed
and compared with experimental data. The puzzle of the neutron-
to proton-induced decay ratio, $\Gamma_n/\Gamma_p$, is addressed
in connection to the two-nucleon induced decay channel and
proposals for more efficient experimental analyses of this ratio
are made. 

\vspace{1cm}

\vspace{0.5cm}


\end{abstract}

\newpage
\tableofcontents

\section{Introduction}

Hypernuclear physics is reaching the stage of a mature science.
Many theoretical and experimental efforts have been devoted to
the subject and a number of comprehensive reviews are already
available
\cite{galR77,povhR78,bandoR85,doverR89,oset90,cohen1,bandoR90,bandoR92,ptp117,gibsonR95,akai97}.
The present review covers only one
aspect of the investigations in the field,
that of the weak
decay of $\Lambda$ hypernuclei. This topic has proved to be a
rich one in the interface between particle and nuclear physics:
both information on particle properties unaccesible with ordinary
elementary reactions, as well as nuclear properties of
hypernuclei, complementary to those from nuclear structure, have
been obtained. Many of the interesting results from hypernuclear
decay have been extensively reported in the reviews quoted above,
hence the present one will be devoted essentially to the new
achievements available from 1990 on.

The early experiments to produce hypernuclei using emulsions or
the $(K^-,\pi^-)$ reaction at CERN and BNL have given room to
many new ones in a large number of experimental facilities:
$\Sigma$ and $\Lambda$ hypernuclei have been produced at KEK
using the $(K^-,\pi)$ reaction with stopped kaons. The same
reaction but with in-flight kaons has been studied at BNL. More
recently, the $(\pi^+,K^+)$ reaction at BNL and KEK has proved to
be very efficient in populating deeply bound $\Lambda$ states
and, when studied away from the forward scattering angles, has
allowed the measurement of asymmetries in the particles emitted
from the decay of polarized hypernuclei. Another source of
information has been obtained from LEAR, where hypernuclei and
their decay are studied using $\bar{p}$ absorption followed by
delayed fission. A similar technique but with protons on nuclei
has been successfully developed at COSY which has allowed the 
measurement of lifetimes of heavy hypernuclei.
The TJNAF laboratory has also joined these efforts and the
$(e,e^\prime K^+)$ reaction is now ready to produce hypernuclei
with a much better energy resolution.
Double strange $\Lambda \Lambda$ and $\Xi$ hypernuclei have also
been obtained from the $(K^-,K^+)$ at BNL, opening new
possibilities to the understanding of the $\Lambda \Lambda$ and
$\Xi N$ interactions as well as to unravel the existence of the
$H$ particle.
Finally, the FINUDA project at DA$\Phi$NE will soon use the
tagged and slow $K^-$ from the $\Phi$ decay into $K^+ K^-$ to
provide data at a rate significantly higher than was available in
the past.

The theoretical developments have run in parallel to the
experiments. The mesonic decay of $\Lambda$ hypernuclei ($\Lambda
\to \pi N$) has confirmed the strong sensitivity to the pion
nucleus optical potential discovered in the past. The total
mesonic decay rate is significantly enhanced due to the pion
interaction
in the nucleus which comes, basically, from the attractive p-wave
part of the optical potential. However, exclusive reactions to a
final closed shell nucleus select basically the repulsive s-wave
part and lead to a reduced partial decay rate. The simultaneous
study of inclusive and exclusive decay reactions can thus be a
good source of information on the pion nucleus interaction,
complementary to the one obtained from pionic atoms and low
energy pion nucleus scattering. The mesonic decay of light
hypernuclei has provided evidence of the strong repulsion of the
$\Lambda N$ interaction at short distances, a property that
follows naturally from available quark models of the strong $YN$
interaction.

The non-mesonic decay mode ($\Lambda N \to NN$) has received much
attention and detailed models which go beyond the original one
pion exchange mechanism are now available, either including the
exchange of several mesons or working explicitly in terms of the
quark degrees of freedom. One of the motivations for developing
these models was the large discrepancies between theory and
experiment in the ratio $\Gamma_n/\Gamma_p$ of neutron- to
proton-stimulated non-mesonic decay, where the one pion model
seems to provide an order of magnitude smaller ratio than the
data. The question is not yet settled but one of the things that
has been recently found is that the consideration of the 
two-nucleon-induced decay mode ($\Lambda N N \to N N N$) enlarges the
error bars from present experimental analyses.
The two-nucleon-induced non-mesonic decay of $\Lambda$
hypernuclei was considered with the hope that it could solve the
$\Gamma_n/\Gamma_p$ puzzle. While it has been seen that this is
not the case, the new decay remains, however, as a channel which
must be taken into account in any attempt of a precise analysis
of the experimental data. Theoretical studies of the neutron and
proton spectra from the decay of $\Lambda$ hypernuclei are now
available and open new doors for a more reliable determination of
the $\Gamma_n/\Gamma_p$ ratio than was possible in the past.

In this review we concentrate only on the weak decays
of $\Lambda$ hypernuclei.
The progress in $\Sigma$ hypernuclei has been scarce both
experimentally and theoretically, however, the narrow $\Sigma$
states claimed by old spectra have not been observed in a recent
experiment carried out at BNL with better statistics. Another
interesting experiment has been the clean measurement of the
$^4_{\Sigma}$He hypernucleus at BNL \cite{nagae98}, 
giving stronger grounds for
the existence and properties of this state than the previous
findings at KEK using stopped kaons. 
Also, the availability of $\Lambda \Lambda$ and $\Xi$
hypernuclei has stimulated many theoretical studies aimed at
obtaining information about the $\Lambda \Lambda$ and $\Xi N$
interaction. A comprehensive recent review on these latter issues
can be found in ref. \cite{akai97}.

\newpage

\section{Weak mesonic decay of hypernuclei}

\subsection{Derivation of the $\Lambda$  width in nuclei}

We make here a formal derivation of the width of $\Lambda$ states
bound
in nuclei. In the first place we evaluate the $\Lambda$ width
for a $\Lambda$ particle moving through infinite nuclear matter.
The
width for finite nuclei is obtained from there using the local
density
approximation (LDA). In a second step we follow a direct approach
to
the evaluation of the $\Lambda$ mesonic width in finite nuclei
and
compare
the methods. Anticipating results we find that the nuclear matter
plus LDA is a good tool to evaluate $\Lambda$
decay widths in nuclei. For the non-mesonic channel, where the energy
carried by the emitted nucleons is large, the LDA
predicts decay rates within 5$\%$ of the finite nuclei results.
The mesonic width is more sensitive to nuclear shell effects given the
little phase space available for the reaction. In medium hypernuclei
around \hyp{12}{C}, 
assuming one uses the correct Q value of the reaction, one may expect
the LDA results to lie within 20--30 \% of the correct answer. 
In lighter hypernuclei the mesonic width approaches gradually the
free one and the uncertainties naturally diminish.
In heavier hypernuclei, where the mesonic rate is reduced in several
orders of magnitude, the errors are larger. In this case the LDA can
only account qualitatively for the mesonic width, however it provides the 
essential features of the reaction, such as the reduction in the rate
with increasing mass of the hypernucleus, as well as the dramatic
increase of the rate due to the pion interaction with the nucleus.

We also discuss here two ways of evaluating the decay process,
one
by means of Feynman diagrammatic technique and the use of meson
and nucleon
propagators, the other one that uses operators and wave functions
and
we show the equivalence of the two methods.

\subsubsection{The propagator method}

The starting point is the $\Lambda \longrightarrow \pi N$
Lagrangian,
accounting for this weak process, which is given by

\begin{equation}
{\cal L}^{\rm W }_{\rm {\scriptscriptstyle \Lambda N} \pi}= -i
G_F \mu^2 \overline{\psi}_{\rm N}
(A_\pi+B_\pi \gamma_5)
\vec{\tau} \vec{\phi}_\pi
\psi_\Lambda \, \left( ^0_1 \right) + {\rm h.c.} \ ,
\label{eq:weak}
\end{equation}
where $G_F \mu^2= 2.21\times 10^{-7}$ is the weak coupling
constant.
The empirical constants
$A_\pi=1.05$ and $B_\pi=-7.15$, adjusted to the observables of
the
free
$\Lambda$ decay, determine
the strength of the parity violating and parity conserving
amplitudes, respectively.
The nucleon, $\Lambda$ and pion fields are given by
$\psi_{\rm N}$, $\psi_\Lambda$ and $\vec{\phi}_\pi$,
respectively, while
the isospin spurion $\left( ^0_1 \right)$ is included
to enforce the empirical
$\Delta I=1/2$ rule,
according to which the strength for the $\Lambda \to \pi^- p$
decay is double than that for $\Lambda \to \pi^0 n$.

A practical way to evaluate the $\Lambda$ width in nuclear matter
and to
introduce  medium corrections is to start from the $\Lambda$
self-energy,
$\Sigma$, associated to the diagram of fig. 1, and then use the
relationship

\begin{equation}
\Gamma = -2 {\rm Im}\, \Sigma .
\end{equation}

\begin{figure}[htb]
       \setlength{\unitlength}{1mm}
       \begin{picture}(80,60)
      \put(42,-22){\epsfxsize=8cm \epsfbox{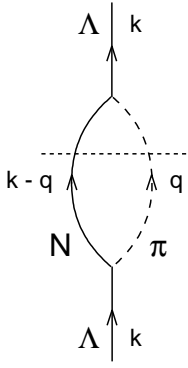}}
        \end{picture}
\caption{
Feynman graph for the free $\Lambda$ self-energy of eq. (3).
The $\Lambda \rightarrow \pi N$ ``cut" is shown (dotted line).
}
\label{fig:repfig1}
\end{figure}

Following standard Feynman rules, and making a non relativistic
reduction of the operators in eq. (1),
the self-energy is readily evaluated as

\begin{equation}
-i \Sigma (k) = 3 (G_F \mu^{2})^{2} \int \frac{d^{4}q}{(2 \pi
)^{4}}
G (k-q)
D(q) \left[ S^{2} + \left( \frac{P}{\mu} \right)^{2} \vec{q}\,^2
\right] \ ,
\end{equation}

\noindent
where $G$ and $D$ are the nucleon and pion propagators,
respectively, $S=A_\pi$, and
$P/\mu = B_\pi/2M$ with $\mu$ and  $M$ the pion and nucleon
masses.

The free nucleon and pion propagators are given respectively by
\begin{eqnarray}
G (k) &=& \frac{1}{k^0 - E (\vec{k}\,) + i \epsilon} \nonumber\\
D (q) &=& \frac{1}{q^{0 \,2} - \vec{q}\,^2 - \mu^2 + i \epsilon}
\ ,
\end{eqnarray}
where the typical nonrelativistic approximations are done in the
nucleon
propagator: only the positive energy part of the nucleon
propagator
is taken and the factor $M/E$ ($E$ being the relativistic energy)
is set equal to unity.

By using the free
nucleon and pion propagators one obtains immediately the free
$\Lambda$ width
\cite{oset90,oset}

\begin{eqnarray}
\Gamma_{\rm free} \equiv \Gamma_{\Lambda} &=& 3 (G_F\mu^{2})^{2}
\int 
\frac{d^{3}q}{(2 \pi )^{3}}
\frac{1}{2 \omega (\vec{q}\,)} 2 \pi \delta (E_{\Lambda} - \omega
(q) - E
(\vec{k}- \vec{q}\,)) \nonumber \\
& & \phantom{3 (G_F\mu^{2})^{2}
\int} \times  \left[ S^{2} + \left( \frac{P}{\mu} \right)^{2}
\vec{q}\,^2 \right] \ ,
\end{eqnarray}

\noindent
where $E_\Lambda, \omega (q), E (\vec{k} - \vec{q}\,) $ are the
energies
of the $\Lambda$, pion and nucleon, respectively.

In a Fermi sea of nucleons, both the nucleon and pion propagators
are changed to
\begin{equation}
G(p)= \frac{1-n(\vec{p}\, )}{p^{0} - E (\vec{p}\, ) - V_{N} + i
\epsilon} +
\frac{n (\vec{p}\, )}{p^{0} - E (\vec{p}\, ) - V_{N} - i
\epsilon}
\end{equation}
\begin{equation}
D(q)= \frac{1}{q^{0\,2} - \vec{q}\,^2 - \mu^{2}
- \Pi (q^{0},{\rm q})} \ ,
\end{equation}
where $V_{N}$ is the nucleon potential, $\Pi (q^{0},{\rm
q}=\mid\vec{q}\,\mid)$ is the
pion
self-energy in the nuclear medium and $n (\vec{p}\, )$ is the
occupation number
 in
the Fermi sea, $n (\vec{p}\, )=1$ for $|\vec{p}\, | \leq k_{F}$
and $n(\vec{p}\, )=0$
for $| \vec{p}\, | > k_{F}$, with $k_{F}$ the Fermi momentum.
An appropriate choice for $V_N$ is the Thomas Fermi potential,
$V_N = - k_F^2/2M$, which becomes $\vec{r}$ dependent when the
local Fermi momentum $k_F (r) = (\frac{3}{2} \pi^2 \rho
(r))^{1/3}$
is used.

The practical way to perform the $q^{0}$ integral in eq. (3) is
through the Wick rotation displayed in fig. 2, where the
analytical
structure of the integrand is shown. The shaded region accounts
for the
discontinuity of the pion propagator due to ${\rm Im}\, \Pi$.
In \cite{oset} one includes particle-hole ($ph$) and
$\Delta$-hole ($\Delta h$)
excitation as a
source of the p-wave pion self-energy plus and extra s-wave
self-energy. Coulomb effects are neglected. Thus
\begin{eqnarray}
\Pi (q) &=& \Pi^{(s)} (q) + \vec{q}\,^2 \tilde{\Pi} (q) \nonumber
\\
{\rm with} ~~~~ \tilde{\Pi} (q) &=&
\frac{\displaystyle\frac{f^2}{\mu^2} U (q)}{1 - g'
\displaystyle\frac{f^2}{\mu^2}
U (q)} \\
{\rm and} ~~~~ U (q) &=& U_N (q) + U_\Delta (q) \ . \nonumber
\end{eqnarray}

\begin{figure}[htb]
       \setlength{\unitlength}{1mm}
       \begin{picture}(90,90)
      \put(30,-30){\epsfxsize=10cm \epsfbox{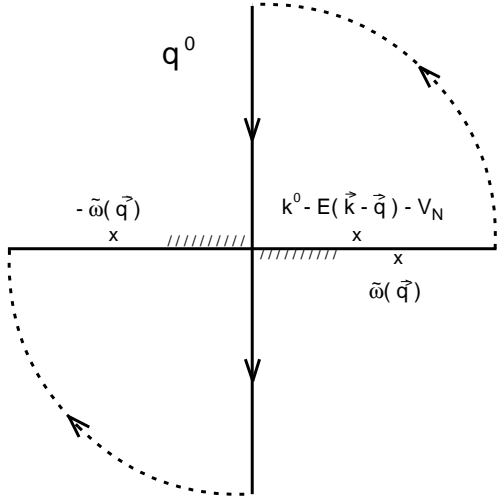}}
        \end{picture}
\caption{
Analytical structure of the integrand of eq. (3) in the complex
$q^{0}$ plane with the nucleon and pion propagators of eqs. (6),
(7).
The renormalized pion propagator pole $\tilde{\omega}(q)$ is
shown. The dashed
lines close to the real axis indicate the analytical cut from
${\rm Im}\, \Pi
(q^{0},q)$ related to the non-mesonic $\Lambda$ decay channel.
}
\label{fig:repfig2}
\end{figure}

In eq. (8), $U (q)$ is the Lindhard function which gets strength
from
$ph$ excitation, $U_N (q)$, and $\Delta h$ excitation,
$U_\Delta (q)$, and $g'$ is the Landau-Migdal
parameter.
In the $\Lambda$ decay case, $U_N (q)$ has both real and
imaginary parts,
the latter one due to those contributions in the integration
over the
internal variables for which the $ph$ excitation appears on the
mass shell.
The function $U_\Delta (q)$ has also real and imaginary parts,
the latter one
coming from the (energy dependent) $\Delta$ width, which is
relatively
small in the $\Lambda$ decay case since the pions emerge with
little energy. Then, for
practical purposes one can neglect ${\rm Im}\, U_\Delta$. In
this case,
the analytical structure of the integrand in eq. (3) with the
medium propagators (6) and (7) shows, in
addition to the cut (region of discontinuity in the real axis due
to ${\rm Im}\, U_N \neq 0$), a pole in $q^0 = \tilde{\omega}
(\vec{q}\,)$
corresponding
 to a renormalized pion energy for which
\begin{equation}
\tilde{\omega} (\vec{q}\,)^{2} - \vec{q}\,^2 -
\mu^{2} - \Pi (\tilde{\omega} (\vec{q}\, ), {\rm q})=0 \ .
\end{equation}
Missing in  fig. 2 is the pole of $G(\vec{k} - \vec{q}\, )$
corresponding to the second term
in eq. (6). This pole lies in the lower half-plane of the figure
and
would
contribute in the Wick rotation only when it happens to be in the
third
quadrant, i.e., $k^{0}-E(\vec{k}- \vec{q}\, )-V_{N} < 0$. But
this corresponds to
$(\vec{k}-\vec{q}\, )$ very large where $n(\vec{k}-\vec{q}\, )=0$
and hence this
term does not contribute.

We come back to the evaluation of $\Sigma (k)$ in the nuclear
medium, eq. (3),
 using the nucleon and pion propagators of eqs. (6) and (7).
In the integral over $q^0$ along the contour of fig. 2, the
contributions of
the arcs at infinity vanish and the integral along the
imaginary
axis contributes only to ${\rm Re}\, \Sigma$, given the property
of the
integrand $I (q^0) = I^* (q^{0*})$. Thus, only the residue of the
integrand at the pole of the first quadrant is responsible for
${\rm Im}\, \Sigma$
and we obtain for the width \cite{oset90,oset}
\begin{eqnarray}
\Gamma (k)&=&-6 (G_F \mu^{2})^{2} \int \frac{d^{3}q}{(2 \pi)^{3}}
[1-n (\vec{k} - \vec{q}\, )] \theta (k^{0} - E (\vec{k} -
\vec{q}\,) - V_{N}) \nonumber \\
&\times & \left[ S^{2} + \left(\frac{P}{\mu}\right)^{2} \vec{q}\,^2
\right]
{\rm Im}\,  \left. \frac{1}{q^{0\,2}
- \vec{q}\,^2 - \mu^{2} - \Pi (q^{0},{\rm q})}
\right|_{q^{0}=k^{0}-E(\vec{k}-\vec{q}\, )-V_{N}} \ .
\end{eqnarray}

In the discussion here we neglect the role of correlations and
form factors,
which are  important only in the non-mesonic decay. We shall
discuss
this in detail when studying this decay mode in Sect. 3.

In eq. (10) we observe the Pauli blocking factor, $1-n$.
Since a $\Lambda$ with momentum $\vec{k}=0$ decays into a nucleon
and pion
with ${\rm q} \simeq
100$ MeV/c, this momentum is smaller than the Fermi momentum
for nuclear
matter density, $ k_{F}= 270$ MeV/c, and
the decay is forbidden by Pauli blocking, i.e,
$1-n(\vec{k}-\vec{q}\,)=0$.
In finite nuclei it is still possible to have mesonic decay since
the $\Lambda$ wave function has some overlap with the nuclear
surface where the Fermi momentum will be smaller than 100 MeV/c.
Moreover, the momentum distribution of the $\Lambda$ leads to
some spreading in the nucleon momenta allowing some of them to
overcome the Pauli blocking.
Nevertheless, the $\Lambda$ mesonic
width
decreases drastically as a function of the mass number.

\begin{figure}[htb]
       \setlength{\unitlength}{1mm}
       \begin{picture}(100,60)
      \put(5,-82){\epsfxsize=14cm \epsfbox{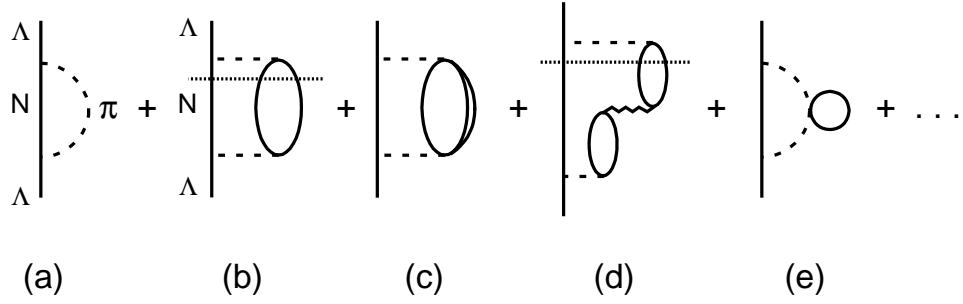}}
        \end{picture}
\caption{
$\Lambda$ self-energy diagrams included in eq. (3) with the
nucleon
and pion propagators of eqs. (6), (7). (a) Free self-energy
graph. (b), (c)
Insertion of p-wave pion self-energy at lowest order. (d)
Generic RPA graph
from the expansion of the pion propagator in powers of the pion
self-energy.
(e) s-wave pion self-energy at lowest order. The cuts represent
the
non-mesonic decay channel.
}
\label{fig:repfig3}
\end{figure}

\begin{figure}[hbt]
       \setlength{\unitlength}{1mm}
       \begin{picture}(80,60)
      \put(25,-27){\epsfxsize=10cm \epsfbox{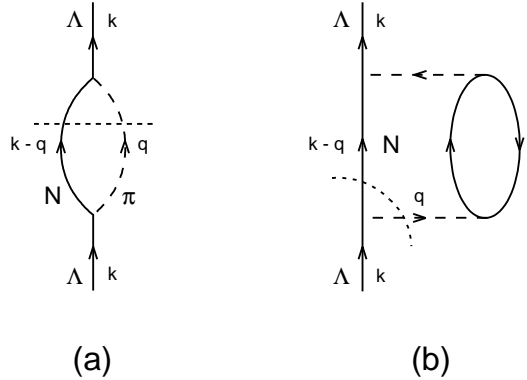}}
        \end{picture}
\caption{
Free and lowest order $\Lambda$ self-energy graph. The dotted
cuts
represent the mesonic decay channel.
}
\label{fig:repfig4}
\end{figure}

The language of propagators used here is very adequate
to provide a unified picture of the $\Lambda$ nuclear
decay. Indeed,
eq. (10) contains not only the modified mesonic channel but also
the
non-mesonic one. This can be seen diagrammatically by expanding
the pion
propagator and taking a $ph$ and $\Delta h$ excitation to account
for the pion
self-energy, $\Pi$. This is depicted in fig. 3. The imaginary
part of a
self-energy diagram is obtained when the set of intermediate
states cut by a
horizontal line are placed simultaneously on shell in the
intermediate
integration. In fig. 3 we observe a source corresponding to
placing on shell
a nucleon and the $ph$ of the pion self-energy. This corresponds
to a channel
where there are no pions and only nucleons in the final state.
The physical
process which has occurred is $\Lambda N \rightarrow NN$, which
corresponds to the standard
non-mesonic channel. Technically, it would be obtained by
substituting in eq.
(10)
\begin{equation}
{\rm Im}\, \frac{1}{q^{0\,2}-\vec{q}\,^2-\mu^{2}
- \Pi}
\longrightarrow \frac{{\rm Im}\, \Pi_{ph}}
{|q^{0\,2}- \vec{q}\,^2 -
\mu^{2} - \Pi |^{2}} \ ,
\end{equation}
where $\Pi_{ph}$ is the pion self-energy due to $ph$
excitations. There is
no overlap between ${\rm Im}\, \Pi_{ph} (q^{0},{\rm q})$
and the pion pole in the propagator
of eq. (10) and thus the separation of the mesonic and
non-mesonic
channels can be done in a clean way.

The mesonic channel would correspond to a different cut, the one
where the
$N$ and the $\pi$ are placed on shell. This is shown
in fig. 4, where the diagram of fig. 4(a) corresponds to the
lowest order in the expansion of the pion propagator. This term,
together with the one in fig. 4(b) and further iterations
contained in eq. (10), lead to a renormalization of the mesonic
width, and an appreciable one, as it was shown
in ref. \cite{oset}.

Technically, the mesonic width can be calculated either by
subtracting the non-mesonic width from the total width, eq. (10),
or, equivalently, by replacing the pion pole
contribution in eq. (5) by the renormalized pion pole given in
eq. (9).
In this latter case the mesonic width is obtained, by analogy to
eq. (5), from
\begin{eqnarray}
\Gamma_{\rm \pi} &=& 3 (G_F\mu^{2})^{2} \int
\frac{d^{3}q}{(2 \pi )^{3}}
\frac{1}{\left|2 \tilde{\omega} (\vec{q}\,)-
\frac{\partial \Pi}{\partial \tilde{\omega}} \right|}
 2 \pi \delta (E_{\Lambda} - \tilde{\omega} (\vec{q}\,) - E
(\vec{k}- \vec{q}\,) - V_N) \nonumber \\
& & \phantom{3 (G_F\mu^{2})^{2} \int} \times  \left[ S^{2} +
\left(\frac{P}{\mu}\right)^{2}
\vec{q}\,^2 \right] \ .
\end{eqnarray}

The qualitative reason for the drastic change on the mesonic
width is given in \cite{oset90,oset}:
the attractive character of the pion self-energy leads
to a larger pion momentum for the same pion energy and thus, to a
larger
nucleon momentum by momentum conservation. Thus, the nucleon has
more chances to
have a momentum bigger than the Fermi momentum, therefore
increasing the
mesonic width.

The width in finite nuclei is obtained in \cite{oset} via the LDA

\begin{equation}
\Gamma = \int d^{3}r| \phi_{\Lambda} (\vec{r} ) |^{2} \Gamma (k,
\rho
(\vec{r} )) \ ,
\end{equation}
where $\phi_{\Lambda}$ is the $\Lambda$ wave function. A further
average
over the momentum distribution of the $\Lambda$ wave function
is also done in \cite{oset}.

\subsubsection{Finite nuclei approach. The
wave
function method}

The finite nuclei approach to the mesonic width was sketched in
\cite{oset86a} and carried out in detail in \cite{iton88}.
The mesonic width is given, in analogy to eq. (5), by
\begin{eqnarray}
\Gamma^{(\alpha)} &=& \frac{1}{2} C^{(\alpha)} (G_F\mu^{2})^{2}
\sum_{N \not\in F}
\int  \frac{d^{3} q}{(2\pi )^{3}} \frac{1}{2 \omega (\vec{q}\,
)} 2\pi \delta (E_{\Lambda} - \omega (\vec{q}\, ) - E_{N})
\nonumber \\
& & \phantom{\frac{1}{2} C^{(\alpha)} (G_F\mu^{2})^{2}
\sum_{N \not\in F}}
\times \left\{ S^{2} \left| \int d^{3} x \phi_{\Lambda} (\vec{x} )
\phi_{\pi}^{(-)} (\vec{q},
\vec{x} )^{*} \phi_{N}^{*} (\vec{x} ) \right|^{2}
\phantom{\left(\frac{P}{\mu}\right)^2} \right. \nonumber \\
& & \phantom{\frac{1}{2} C^{(\alpha)} (G_F\mu^{2})^{2}
\sum_{N \not\in F}
\int}
+ \left. \left(\frac{P}{\mu}\right)^{2} \left| \int d^{3} x
\phi_{\Lambda} (\vec{x} )
\vec{\nabla}
\phi_{\pi} ^{(-)} (\vec{q},\vec{x} )^{*} \phi_{N}^{\, *}
(\vec{x})\right|^{2} \right\} \ ,
\end{eqnarray}
where $\phi_{N}$ is the wave function of the nucleon states and
$\phi_{\pi}^{(-)*}$ corresponds to an outgoing solution of the
Klein Gordon
equation normalized to a plane wave asymptotically $({\rm e}^{-i
\vec{q} \vec{x}})$.
The index $\alpha$ stands now for $\pi^{-} p$ or $\pi^{0} n$
decay,
with $C^{(p)} = 4, C^{(n)} =2$, which one separates here since,
due to shell
effects, these channels can depart drastically from the
elementary
$\Delta I = 1/2$ rule.

The sum in eq. (14) runs only over non occupied nucleon states in
the shell
model. On the other hand,
$\phi_{\pi}^{(-)*}$ is a
solution of the Klein Gordon equation with a proper optical
potential (or
pion self-energy, $\Pi = 2 \omega V_{opt}$), including Coulomb
effects through $V_{c}$, i.e.,

\begin{equation}
\left[- \vec{\nabla}^{2} + \mu^{2} + 2 \omega V_{opt}
(\vec{x})\right]
\phi_{\pi}^{(-)}
(\vec{q},\vec{x})^{*}= \left[\omega - V_{c} (\vec{x})\right]^{2}
\phi_{\pi}^{(-)}
(\vec{q}, \vec{x})^{*} \ .
\end{equation}
The effects of using $\phi_\pi^{(-)*}$ instead of a plane wave
are rather drastic and increase the
mesonic width in
about two orders of magnitude in heavy nuclei
\cite{iton88,moto88,moto92},
in qualitative
agreement with the nuclear matter results of ref. \cite{oset}.

The arguments for the renormalization are expressed now in the
alternative
language as follows: the attraction caused by the pion
self-energy leads to higher momentum components in the pion wave
function and, as a consequence, the
matrix element of eq. (14), which involves
the $\Lambda$ wave function in the 1s$_{1/2}$ ground state and a
nucleon wave function in an unoccupied orbit, is considerably
enhanced.
To see this let us write the pion wave function in terms of its
Fourier transform
$$
\phi_\pi^{(-)}(\vec{q},\vec{x})^* = \int d^3q^\prime {\rm e}^{- i
\vec{q}\,^\prime \vec{x}}
\tilde{\phi}_\pi(\vec{q},\vec{q}\,^\prime\,) \ ,
$$
and recall that for $\mid \vec{x} \mid \to \infty$ it behaves as
${\rm e}^{-i \vec{q} \vec{x} }$. For values of $\vec{x}$ inside
the nucleus, due to the attractive character of the pion nucleus
potential, the pion wave function builds up $\vec{q}\,^\prime$
momentum components such that $\mid \vec{q}\,^\prime \mid >
\mid \vec{q}\, \mid$.
Then, the strength of the transition,
$\mid \langle N \mid {\rm e}^{-i\vec{q}\,^\prime \vec{x}} \simeq
1 - i \vec{q}\,^\prime \vec{x} - (\vec{q}\,^\prime \vec{x})^2/2 +
\dots  \mid \Lambda \rangle \mid$,
appearing in eq. (14), is governed by the momentum
$\vec{q}\,^\prime$ which, being larger in magnitude than
$\vec{q}$, leads to enhanced matrix elements with respect to
those using the asymptotic free pion wave.
In the two languages the physical consequences are the same: an
increased
probability of reaching the unoccupied states and thus an
enhancement of the
mesonic width.

\subsubsection{Equivalence of the propagator and wave function
methods}

Even if apparently the two methods discussed above
look quite different, the physics contained in the
propagator and wave function methods is equivalent. A
formal
derivation of the connection between the methods can be found in
\cite{Os94}, which we reproduce here for completeness,
complementing some points concerning the pion propagator in
infinite nuclear matter and finite nuclei .

Let us start from the pion propagator in finite nuclei written in
coordinate space
\begin{equation}
D(\vec{x},\vec{y},\omega ) = \sum_{n}
\frac{\phi_{n} (\vec{x}) \phi^{*}_{n} (\vec{y})}{\omega^{2}
- \epsilon_{n}^{2} + i \eta} \ ,
\end{equation}
where $\phi_{n} (\vec{x})$ are the pion wave functions in the
nucleus and
$\epsilon_{n}$ their corresponding energies. Ignoring pionic
bound states,
which do not play a role in our problem, we can identify the
pionic wave
functions by the asymptotic momentum $\vec{q}$. Hence their
energy is given by
$\omega (\vec{q}\,) = (\vec{q}^{\, 2} + \mu^{2})^{1/2}$. The sum
over the index
$n$ is then replaced by an integral over $\vec{q}$ as given below

\begin{equation}
D_{\pi}(\vec{x}_{1},\vec{x}_{2};E_{\pi})= \int
\frac{d^{3}q}{(2\pi )^{3}} \frac{\phi_{\pi} (\vec{q},\vec{x}_{1})
\phi_{\pi}^{*}
(\vec{q}, \vec{x}_{2})}{E^{2}_{\pi} - \omega (\vec{q}\,)^{2} + i
\eta}  \ .
\end{equation}

For simplicity in the derivation we shall take the parity
violating part of the width
(the one providing the largest contribution to the
mesonic decay) and will not distinguish
between $\pi^0$ or $\pi^-$ decay. Hence, from eq. (14) we obtain
\begin{eqnarray}
\Gamma_{S} &=& 3(G_F \mu^{2})^{2} S^{2} \sum_{N \not\in F} \int
\frac{d^{3}q}
{(2 \pi )^{3}}\frac{1}{2\omega (\vec{q}\,)} 2 \pi \delta
(E_{\Lambda} - E_{N} -
\omega (\vec{q}\,)) \nonumber \\
&& \phantom{3(G_F \mu^{2})^{2} S^{2} \sum_{N \not\in F}}
\times \left| \int d^{3} x \phi_{\Lambda} (\vec{x}) \phi_{\pi}^{*}
(\vec{q},\vec{x}) \phi^{*}_{N} (\vec{x})\right|^{2} \ ,
\end{eqnarray}
which can be rewritten as
\begin{eqnarray}
\Gamma_{S} &=& 3(G_F\mu^{2})^{2} S^{2} \int d^{3}x_{1} \int d^{3}
x_{2}
\phi_{\Lambda}^{*} (\vec{x}_{1}) \phi_{\Lambda} (\vec{x}_{2})
\phi_{N} (\vec{x}_{1}) \phi_{N}^{*} (\vec{x}_{2}) \nonumber \\
&\times &\int
\frac{d^{3} q}
{(2 \pi )^{3}} \frac{1}{2 \omega (\vec{q}\,)} 2 \pi \delta
(E_{\Lambda} -
E_{N} - \omega (\vec{q}\,))
\phi_{\pi} (\vec{q},\vec{x}_{1}) \phi_{\pi}^{*}
(\vec{q},\vec{x}_{2}) \ ,
\end{eqnarray}
or, by virtue of eq. (17), as
\begin{eqnarray}
\Gamma_{S} &=& 3 (G_F \mu^{2})^{2} S^{2} \sum_{N \not\in F} \int
 d^{3} x_{1} d^{3}
x_{2} \phi_{\Lambda}^{*} (\vec{x}_{1}) \phi_{N}
(\vec{x}_{1}) (-2)
{\rm Im}\, D_{\pi}
(\vec{x}_{1}, \vec{x}_{2}; E_{\pi} = E_{\Lambda} - E_{N})
\nonumber \\
& & \phantom{ 3 (G_F \mu^{2})^{2} S^{2} \sum_{N \not\in F} }
\times \theta
(E_{\Lambda}
- E_{N}) \phi_{\Lambda}(\vec{x}_{2}) \phi_{N}^{*} (\vec{x}_{2})
\ .
\end{eqnarray}
Now, in order to connect with eqs. (10) and (13) one makes a
local density
approximation. In the first step one evaluates $\Gamma$ for a
slab of
infinite nuclear
matter and in the second step one replaces the
width in the infinite slab by an integral
over the nuclear volume assuming slabs of matter in each $d^{3}
r$ of the
nucleus with local density $\rho (\vec{r})$ and with a
probability of finding
the $\Lambda$ particle given by $|\phi_{\Lambda} (\vec{r})|^{2}$.
This last
step is implemented by means of eq. (13). Hence we should see how
we reproduce
now eq. (10) when we assume in eq. (20) a slab of infinite
nuclear matter. For
this purpose we have to replace for the nucleon sector

$$
\sum_{N \not\in F} \rightarrow V \int \frac{d^{3} p}{(2 \pi)^{3}}
(1 - n (\vec{p}))
$$

$$
\phi_{N} (\vec{x}) \rightarrow \frac{1}{\sqrt{V}}{\rm e}^{i
\vec{p}
\vec{x}}
$$

\begin{equation}
E_{N} \rightarrow E(\vec{p}) - V_{N} \ ,
\end{equation}

\noindent
and for the $\Lambda$ wave function

\begin{equation}
\phi_{\Lambda} (\vec{x}) \rightarrow \frac{1}{\sqrt{V}}
{\rm e}^{i\vec{k} \vec{x}} \ .
\end{equation}
Now, in the infinite slab of nuclear matter the pion propagator
of eq. (17) is replaced by
\begin{equation}
D_{\pi} (\vec{x}_{1},\vec{x}_{2},E_{\pi}) \rightarrow \int
\frac{d^{3}q}{(2
\pi )^{3}} \frac{{\rm e}^{i\vec{q}(\vec{x}_{1}
-\vec{x}_{2})}}{E^{2}_{\pi} - \omega
(\vec{q}\,)^{2} - \Pi (E_{\pi} , \vec{q}\,)} \ ,
\end{equation}
where $\Pi (E_{\pi}, \vec{q}\,)$ is the pion self-energy, which
is a function
of $\rho$. Note that for values of $\vec{x}_{1},\vec{x}_{2}$ far
away from the
nucleus, eqs. (17) and (23) are equivalent since there $\rho = 0$
and $\Pi$
(in the LDA) will be zero. At other densities,
$\Pi$ will be different from zero and the integral of eq. (23)
gives rise to
other momentum components, modulating the plane wave in the
numerator and
providing a kind of WKB approximation to the wave functions
in the
numerator of eq. (17). The LDA gives rise to a
variable local momentum and hence a distorted pion wave.

By substituting eqs. (21), (22), (23) in eq. (20) we obtain:
\begin{eqnarray}
\Gamma_{S} &=& -6 (G_F \mu^{2})^{2} S^{2} \int \frac{d^{3}p}{(2
\pi)^{3}}
\int \left. \frac{d^{3}q}{(2\pi )^{3}} (1-n(\vec{p}))
{\rm Im}\, D_{\pi} (q) \theta (q^{0})
\right|_{q^{0} = E_{\Lambda} - E (\vec{k} - \vec{q}\,) - V_{N}}
\nonumber \\
& & \phantom{ -6 (G_F \mu^{2})^{2} S^{2} \int}\times \int d^{3}
x_{1}
d^{3} x_{2} \frac{1}{V} {\rm e}^{i \vec{p}(\vec{x}_{1} -
\vec{x}_{2})}
{\rm e}^{i\vec{q}(\vec{x}_{1} - \vec{x}_{2})}{\rm e}^{-i\vec{k}
(\vec{x}_{1}
- \vec{x}_{2})} \ ,
\end{eqnarray}
with $D_{\pi} (q)$ given by eq. (7). Finally, by means of the
relationship
$(2 \pi )^{3} \delta^{3} (0) = \int d^{3} x = V$ we can cast eq.
(24) as

\begin{equation}
\Gamma_{S}= -6 (G_F \mu^{2})^{2} S^{2} \int \left. \frac{d^{3}
q}{(2\pi )^{3}} (1 - n(\vec{k} - \vec{q}\,))
{\rm Im}\, D_{\pi} (q) \theta (q^{0})\right|_{q^{0}= E_{\Lambda}
-E (\vec{k}-\vec{q}\,) - V_{N}} \ ,
\end{equation}
which coincides with the s-wave contribution to $\Gamma$ from
eq. (10). This
establishes the equivalence between the finite nuclei wave
functions method and the propagator method, the latter one
complemented with the LDA.

A further consideration can be done regarding the pion
propagators
in finite nuclei and nuclear matter, eqs. (17) and (23)
respectively.
Neglecting, for simplicity, the Coulomb potential, assuming a
smooth energy dependence of the pion self-energy and taking $\Pi$
for a value of $E_\pi$ around the pion pole, the Klein-Gordon
equation leads to
\begin{eqnarray}
\left[- \vec{\nabla}\, ^2 \right. &+& \left. \mu^2 + \Pi (q, \rho
(\vec{x}_2)) -
E_\pi^2\right] \, D^{FN}_\pi (\vec{x}_1, \vec{x}_2, E_\pi) =
\nonumber \\
& &  \int \frac{d^3 q}{(2 \pi)^3} \frac{\omega (\vec{q}\,)^2 -
E_\pi^2}{
E_\pi^2 - \omega (\vec{q}\,)^2} \phi_\pi (\vec{q}, \vec{x}_1)
\phi^*_\pi (\vec{q}, \vec{x}_2) = - \delta^3 (\vec{x}_1 -
\vec{x}_2) \ ,
\end{eqnarray}
where in the last step we use the fact that the pionic wave
functions
form a complete set of states.

We can proceed equally with the nuclear matter propagator
and we obtain
\begin{eqnarray}
\left[- \vec{\nabla}\,^2 \right. &+& \left. \mu^2 + \Pi (q, \rho
(\vec{x}_2)) -
E_\pi^2 \right] \, D_\pi^{NM} (\vec{x}_1, \vec{x}_2, E_\pi) =
\nonumber \\
& & \int \frac{d^3 q}{(2 \pi)^3 } \frac{\vec{q}\,^2 + \mu^2
 + \Pi (q , \rho
(\vec{x}_2)) - E_\pi^2}{E_\pi^2 - \vec{q}\,^2 - \mu^2 - \Pi
(q,\rho (\vec{x}_2))} {\rm e}^{i \vec{q} (\vec{x}_1 - \vec{x}_2)}
=
- \delta^3 (\vec{x}_1 - \vec{x}_2) \ .
\end{eqnarray}

The integral over $\vec{q}$ gives the same results in eqs. (26)
and
(27), leading to a local function, and providing further insight
in
the meaning of the LDA approximation in inclusive
processes.

Note also that for a fixed $E_\pi$ of the pion, the value of
${\rm q}$
at the pole of the two propagators is different, but also its
meaning. In the case of finite  nuclei we have $E_\pi = \omega
(\vec{q}\,)$ and the value of $\vec{q}$
is the asymptotic one when the pion leaves the nucleus. In the
case
of infinite nuclear matter we have $E_\pi^2 =
\omega (\vec{q}\,)^2 + \Pi (E_\pi, q, \rho (\vec{r}))$ which
gives

\begin{equation}
E_\pi \simeq \omega (\vec{q}\,) + \frac{\Pi (r)}{2 \omega
(\vec{q}\,)} \equiv
\omega (\vec{q}\,) + V_{opt} (r) \ .
\end{equation}
Hence, in this latter case, the value of ${\rm q}$ is the local
momentum
of the pion such that the total energy, kinetic plus potential,
is
the asymptotic energy of the pion.

\subsection{The mesonic width and the occupation number}

We have seen that Pauli blocking is the major factor in reducing
the
mesonic
width of heavy $\Lambda$ hypernuclei. It was suggested that
because real
interacting nuclei have the ``occupied'' states partly
unoccupied, the mesonic
width should be enhanced with respect to a calculation with fully
occupied states up to the
Fermi level \cite{bando85}. In the nuclear matter approach of
Sect. 2.1.1 this is
easily visualized by recalling a realistic picture of the
occupation number
of the Fermi sea \cite{fanto84},
which is depicted in fig. 5. For the states below
the Fermi momentum the level of occupancy is of the order of $85
\%$ and, by
assuming that in the $\Lambda$ decay the nucleons can occupy the
$15 \%$
vacancy of these states, we would guess that the mesonic width
would stabilize
at the level of about $10 \%$ of the free width for heavy nuclei
(taking into
account the absorption of pions
on their way out of the nucleus). If this were the
case the mesonic width could serve as a measure of the occupation
number in
the Fermi sea. The argument is very appealing and intuitive,
however, it is
incorrect and leads to an overestimate of the width by about
three orders of
magnitude in heavy nuclei.

\begin{figure}[htb]
       \setlength{\unitlength}{1mm}
       \begin{picture}(100,70)
      \put(5,-45){\epsfxsize=13cm \epsfbox{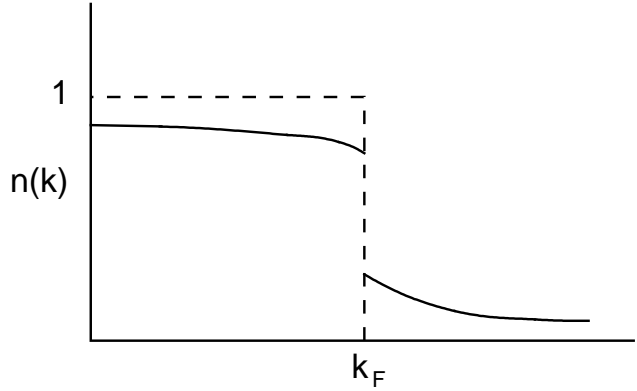}}
        \end{picture}
\caption{
Schematic representation of the nucleon occupation number for an
interacting Fermi sea.
}
\label{fig:repfig5}
\end{figure}

The detailed discussion of this problem was done in ref.
\cite{pedro91}.
The falacy
in the argumentation lies in the the following: since the Fermi
distribution
$n (\vec{p})$ of eq. (6) is only a first approximation to the
momentum
distribution, it looks like a sensible improvement to substitute
the
nucleon Fermi propagator of eq. (6) by
\begin{equation}
\frac{1 - n_{1} (\vec{k})}{k^{0} - E(\vec{k}) + i \epsilon } +
\frac{n_{1} (\vec{k})}{k^{0} - E(\vec{k}) - i \epsilon } \ ,
\end{equation}
where $n_{1} (\vec{k})$ is the realistic occupation number in
nuclear matter and is then $\neq 0$ for all values of $\vec{k}$.
This is actually and approximation often found in the literature,
particularly in works of condensed matter and heavy ions. Given
the analytical structure of the propagator, eq. (29), identical
to
the one of eq. (6), the formal result for the width is the same
as
in eq. (10) replacing only $1 - n (\vec{k} - \vec{q}\,)$
by $1 - n_1 (\vec{k} - \vec{q}\,)$, in which case we would find
widths of
the order of 15$\%$ of the free width. However, eq. (29) is not
an improvement over the propagator of eq. (6), but quite the
opposite, as we shall see.
 The realistic nucleon propagator for an interacting Fermi sea is
given in terms of the spectral
functions by
\begin{equation}
G(k^{0},k) = \int_{-\infty}^{\mu_p} d\omega \frac{S_{h} (\omega ,
k)}
{k^{0} - \omega - i \epsilon} + \int_{\mu_p}^{\infty} d\omega
\frac{S_{p} (\omega , k)}{k^{0} - \omega + i \epsilon} \ ,
\end{equation}
with $\mu_p$ the chemical potential.

When performing the calculations of the mesonic width with this
nucleon propagator one obtains the factor

\begin{equation}
\int_{\mu_p}^{\infty} d\omega S_{p} (\omega , \vec{k}- \vec{q}\,)
2 \pi \delta
(k^{0} - \omega - \omega (\vec{q}\,))
\end{equation}
replacing the factor
\begin{equation}
[1-n (\vec{k} - \vec{q}\, )] 2 \pi \delta (k^{0} - E (\vec{k} -
\vec{q}\,) -
\omega (\vec{q}\,))
\end{equation}
in eq. (5), when the Pauli blocking factor of eq. (10) is
implemented. Eqs.
(31) and (32) bare some intuitive resemblance because

\begin{equation}
\int_{\mu_p}^{\infty} d\omega S_{p} (\omega, \vec{k} - \vec{q}\,)
= 1 - \int^{\mu_p}_{-\infty} S_{h} (\omega , \vec{k}-\vec{q}\,)
d \omega = 1 - n_{1}(\vec{k} - \vec{q}\, ) \ .
\end{equation}
However, the presence of the $\delta$ function in eq. (31)
prevents from factoring out the integral of $S_p$ shown in eq.
(33). Furthermore, because of restrictions of the phase
space (energy
and momentum conservation) the range of values of $\omega$
allowed are very
small compared to the range $(\mu_p , \infty )$ needed in eq.
(33) to obtain
$1 - n_{1} (\vec{k} - \vec{q}\,)$
of the interacting Fermi sea. In physical terms we can
interprete it in the following way: the occupation number $n_{1}
(\vec{k} -
\vec{q}\, )$ is an integral for all the energies of the nucleon,
$\omega$, of
the probability of finding a nucleon with
momentum $\vec{k} - \vec{q}$ and an
energy $\omega$, which is given by the spectral function $S_{h}
(\omega ,
\vec{k} - \vec{q}\, )$. However, in a physical decay process we
have conservation
of energy and momentum and hence there are severe restrictions to
the values
of the energies that the nucleon can have. This is why the
occupation number
$n_{1} (\vec{k} - \vec{q}\,)$ cannot be factored out.

The actual calculations carried out in ref. \cite{pedro91} showed
that for light
and medium nuclei the use of the spectral representation for the
nucleon
propagator, eq. (30), instead of the one of the noninteracting
Fermi sea,
eq. (6), has negligible consequences in the mesonic width (of the
order of
$6\%$ corrections in \hyp{16}{O}).
The corrections can be of the order of $50\%$ in heavy
nuclei, but in all cases, when the pionic renormalization is
taken into
account, one can disregard these effects.

These findings have been of relevance in showing similar problems
in the
study of other physical processes, like in the contribution of
the pion cloud
to $K^{+}$ nucleus scattering where it is shown \cite{garcia95}
that one cannot
relate the effect of the pion cloud to the pion excess number in
the nucleus
as assumed in refs. \cite{akul92,jiang92}.
Similar spectacular differences between the use of the occupation
number  and the spectral function are seen in the study of the
nuclear structure functions of deep inelastic scattering at
values of $x >1$ \cite{Fm96}. For values of $x = 1.3$ the use of
the
occupation number gives values of the structure function two
orders
of magnitude bigger than the results with the spectral functions,
which
agree with experiment \cite{Be94}.

\subsection{Results for the mesonic width}

We separate here three regions of heavy, medium and light nuclei,
where different physical phenomena are explored by means of the
mesonic decay.

Experimental data for light and medium nuclei of the last decade
can
be found in \cite{grace85,saka89,bar86,szymanski,saka91}.

\subsubsection{Medium and heavy nuclei}

In refs. \cite{iton88,moto88,moto92} one can find abundant
results in different nuclei which
are rather realistic. These results have been recently improved
\cite{niev93a}
by
a more accurate description of the energy balance in the
particular
reactions, taking into account transitions to the bound and
continuum nucleon
states and using a pion nucleus optical potential which has been
derived
theoretically and leads to a good description of pionic atoms
data and of elastic, reaction and absorption cross sections in
the
scattering processes
\cite{niev93b}. The imaginary part of the pion nucleus optical potential
is split into two terms related
to pion absorption and quasielastic scattering. In
\cite{niev93a}
 the
pion quasielastic events are not removed from the pion flux, as
it corresponds
to the actual experimental observation, while the use of a full
distortion of
the pion with the total optical potential, as done in
\cite{iton88,moto88,moto92},
inevitably
removes the pion quasielastic events, together with the pion
absorption
events.
Though conceptually important, this refinement turns out to be of
little
practical relevance in the present problem given the small energy
that the
pions carry and the very small phase space for quasielastic
collisions
\cite{niev93a}.
However, other considerations, particularly the energy balance in
the reactions, makes the widths in heavy nuclei for
$\pi^{-}$-decay about
one order of magnitude smaller than those of ref. \cite{moto92}.

\begin{figure}[htb]
       \setlength{\unitlength}{1mm}
       \begin{picture}(100,120)
      \put(35,0){\epsfxsize=8cm \epsfbox{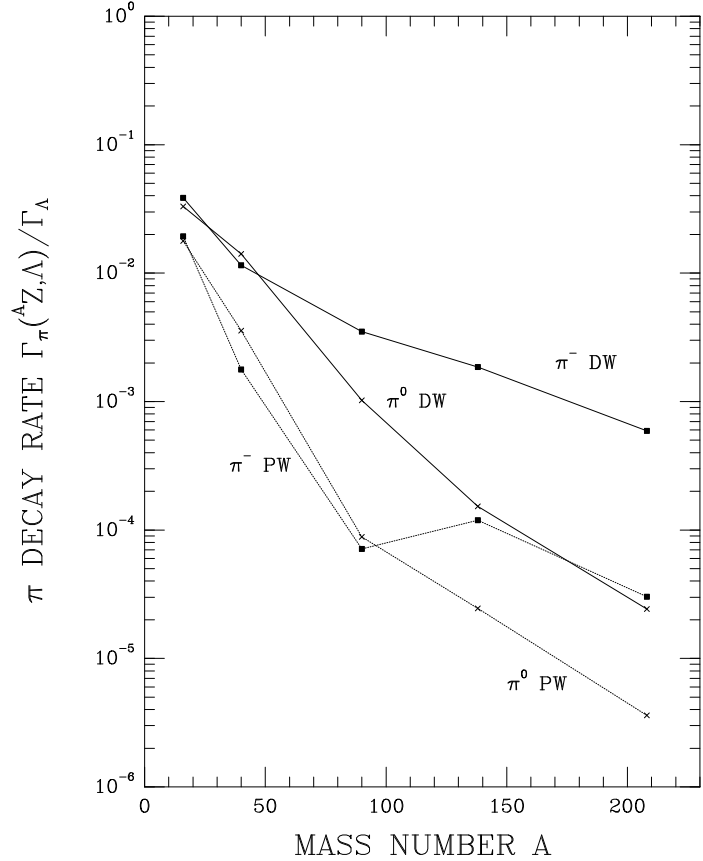}}
        \end{picture}
\caption{
Pionic decay rate for $\pi^0$ and $\pi^{-}$ as a function of
the
mass number (of the host nucleus, $^{16}$O, $^{40}$Ca, $^{90}$Zr,
$^{138}$Ba,
and $^{208}$Pb). The two lower lines show the calculation with
plane waves
for the pion and the two upper lines the results with pion
distorted waves.
}
\label{fig:mesonic}
\end{figure}

In fig. 6 we show the prediction of ref. \cite{niev93a}
for different nuclei and for
$\pi^{0}$ and $\pi^{-}$ decay, with plane waves and the
renormalized
pion wave function. The drastic effects of the pion
renormalization are seen
there and are a bit smaller than in former works because the
energy balance
forces the pions to come out with smaller energies than in the
previous approaches
and the attractive effects of the p-wave part of the optical
potential are
then diminished.
Shell effects and energy balance are important here because of
the small energies left for the pions. They are responsible for the
slight increase of the $\Lambda$ width in $^{139}_\Lambda$Ba
with respect to $^{91}_\Lambda$Zr seen in the figure for the 
plane wave calculation. When the 
interaction of the pions with the
nucleus is considered, the extra attraction
felt by the pions gives more room for pionic decay
such that the nuclear energy differences in the
energy balance are not so crucial.

\subsubsection{Shell effects in medium nuclei}

The energy balance for some nuclei makes the results somewhat
surprising.
Of particular relevance are the results in $_{\Lambda}^{12}$C
shown in Table \ref{tab:mesonic1}.

\begin{table}[hbt]
\centering
\caption{Pionic partial decay rates in the decay of \hyp{12}{C}}
\begin{tabular}{|l|l|l|l|}
\hline
  & $\Gamma_{\pi^{0}}/\Gamma_{\Lambda}$ &
$\Gamma_{\pi^{-}}/\Gamma_{\Lambda}$
 & $\Gamma_{\pi^{0}}/\Gamma_{\pi^{-}}$\\
\hline
\cite{niev93a} & 0.159 & 0.086 & 1.86\\
\cite{moto92} & 0.13 & 0.098 & 1.32\\
\hline
EXP: \cite{szymanski} & & $0.052_{-0.035}^{+0.063}$  & \\
EXP: \cite{saka91} & $0.217 \pm 0.084$  &  & \\
\hline
\end{tabular}
\label{tab:mesonic1}
\end{table}

In ref. \cite{moto88}, using different pion optical potentials, the authors  
quote 0.169 and 0.126 for $\Gamma_{\pi^0}/\Gamma_\Lambda$ and
0.134 and 0.107 for $\Gamma_{\pi^-}/\Gamma_\Lambda$.
Although with large errors, the experimental results confirm
these striking
theoretical predictions which show a large departure from the
$\Delta I = 1/2$ rule in nuclei. Indeed, we would naively expect
 $\Gamma_{\pi^{0}}/\Gamma_{\pi^{-}} \sim 0.5$  (this should be the case for 
$N=Z$ core nuclei), while it is of the order of 2 for the
results of ref. \cite{niev93a} 
and slightly larger than one for the
results of refs. \cite{moto88,moto92}. This is
mostly due to nuclear shell effects because of the large
differences
between the $Q$ values met in these reactions.

\subsubsection{Light nuclei: Short range repulsion and quark
models}

Another interesting finding is seen in very light nuclei. The
mesonic width of
$_{\Lambda}^{5}$He has attracted particular attention. There, in
addition to
the pion renormalization, the repulsive character of the $\Lambda
N$
interaction and the relatively weaker medium range attraction,
compared to the
$NN$ interaction, push the $\Lambda$ wave function
to the surface
of the nucleus, weakening the Pauli blocking effect and thus
enhancing the
mesonic decay \cite{kuri85,oset86b}.
The experimental numbers clearly favour potentials
with a repulsive $\Lambda N$ core. One should note that such a
repulsion
automatically appears in quark based models of the $\Lambda N$
interaction \cite{straub90}.
The study of the $^{5}_{\Lambda}$He decay using a quark model
based
hypernuclear wave function \cite{straub} leads to the results 
shown in
Table \ref{tab:mesonic2}, where they are compared to
those of \cite{moto91}, which uses a $\Lambda$
wave function from the modified YNG $\Lambda N$ 
interaction \cite{yama85} with a strong short range repulsion,
and to those of
\cite{Kf95} for the Isle $\Lambda$-nucleus repulsive potential
at short distances \cite{Ku82}.

\begin{table}[htb]
\centering
\caption{Pionic partial decay rates in the decay of \hyp{5}{He}}
\begin{tabular}{|l|l|l|l|}
\hline
 & $\Gamma_{\pi^{-}} / \Gamma_{\Lambda}$ & $\Gamma_{\pi^{0}}
/ \Gamma_{\Lambda}$
& $\Gamma_{\pi} / \Gamma_{\Lambda}$\\
\hline
\cite{straub}& 0.431 & 0.239 & 0.670\\
\cite{moto91} & 0.393 & 0.215 & 0.547\\
\cite{Kf95} & 0.40 & 0.20 & 0.5\\
\hline
EXP: \cite{szymanski} & $0.44 \pm 0.11$  &  $0.18 \pm 0.20$ &
$0.59_{-0.31}^{+0.44}$ \\
\hline
\end{tabular}
\label{tab:mesonic2}
\end{table}

There is a fair agreement of the theoretical results 
\cite{straub,moto91,Kf95} with experiment \cite{szymanski}, account taken
of the large experimental errors.

Although the effect of the renormalization of the pion in the
medium leads to an enhancement of the inclusive mesonic rate, it
is possible to find particular decay channels where the effect is
reversed. This has been shown in ref. \cite{neelima97} for the
decay $_\Lambda^4{\rm H} \to ^4{\rm He} + \pi^-$. The final
nucleus $^4$He selects the s-wave part of the pion optical
potential, which is repulsive, hence leading to a reduction of
the rate with respect to a calculation using free pion waves.
This also means that, through the study of the mesonic decay in
selected channels, one can learn about different parts of the
pion nucleus optical potential.

The mesonic decay of the hypertriton, \hyp{3}{H}, has also been
studied recently \cite{kam97}, using wave functions for the
hypertriton, $3N$ bound and $3N$ scattering states which are
solutions of the Faddeev equations. The $^3{\rm He}+\pi^-$,
$p+d+\pi^-$ and $p+p+n+\pi^-$ decay channels and the
corresponding ones for $\pi^0$ have been studied. The total
mesonic width is 92\% of the free $\Lambda$ width and compares
favourably with the experimental data for the hypertriton
lifetime, which ranges between $(2.20^{+1.02}_{-0.53})\times
10^{-10}$
sec to $(2.64^{+0.92}_{-0.54})\times 10^{-10}$ sec \cite{key73}.

An interesting problem related with the decay of light
hypernuclei is the ratio $\pi^+/\pi^-$ in the decay of
\hyp{4}{He}. This ratio is measured to be about 5\% and cannot
come from the direct decay of the $\Lambda$ ($\Lambda \to \pi^-
p$, $\pi^0 n$). The problem concentrated some attention in the
past \cite{dal64}, where rescattering of the $\pi^0$, $\pi^0 + p
\to \pi^+ + n$, or $\Sigma^+$ decay, $\Sigma^+ \to \pi^+ n$,
following the $\Lambda p \to \Sigma^+ n$ conversion were
considered. However, the results gave a fraction much too small
compared to experiment. The problem was revived recently
\cite{CG97}, using updated information, and a result for that
fraction of 1.2\% was obtained, which, although being a factor
of two larger than in \cite{dal64}, is still too small compared
with the experimental value.

A more recent work \cite{GT97} proposes a solution which consists
of adding to the \hyp{4}{He} wave function some
$\Sigma^+ + {}^3$H component, which is relatively sizable in
view of the strong $\Lambda N \to \Sigma N$ conversion. The
$\Sigma^+$ would then decay into $\Sigma^+ N \to \pi^+ n N$,
which leads to a s-wave spectrum as observed experimentally
\cite{key76}.

On the other hand, a different view is taken in \cite{oka97},
where it is suggested that the $\Delta I=3/2$ component of the
weak interaction is the responsible for the relatively high
fraction of $\pi^+$ emission.

\subsection{Conclusions on mesonic $\Lambda$ decay}

We have made a review of the present situation concerning the
mesonic decay of
$\Lambda$ hypernuclei. We have established the formal link
between the
propagator method, where the huge enhancement of the pionic decay
width was
first reported, and the finite nuclei approach with wave
functions and matrix
elements. Shell effects and precise values of the nuclear binding
energies are
also important in the mesonic width and they are best taken into
account in
the finite nuclei approach. The intuitive and appealing, but
falacious, link
between the nucleon occupation number and the mesonic width has
also been
discussed, which has served to unveil rough approximations used
in other
processes to link the pion excess number with contributions of
the nuclear
meson cloud to some physical observables, like $K^{+}$ nucleus
scattering
or deep inelastic scattering.
We have also discussed the relevance of the short range $\Lambda
N$ repulsion
in the mesonic width of light hypernuclei and showed how 
different models, all of them accounting for this repulsion, can
provide a fair description of the experimental data. We should
note in this respect that microscopical quark models lead naturally 
to such kind of repulsion, although a description in terms of
the more conventional meson exchange picture is also posible.

With the limited amount of experimental data available on the
mesonic
channel, the
amount of physical information obtained is remarkable. As
just mentioned, there is
support for
strong $\Lambda N$ repulsion at short distances.
The process
also provides us with
the most striking renormalization effect due to the pion nucleus
interaction.
 The sensitivity of the
$\Lambda$ decay to the pion nucleus optical potential can  serve
as a
tool to choose between different theoretical descriptions of the
complex
mechanisms of pion nucleus interaction. The decay channel into
$\pi^{0}$
can be an excellent instrument to learn about $\pi^{0}$ nucleus
interaction, and so on.

It is clear that a systematic experimental search in many nuclei
of the
mesonic decay channel  will provide
us with very valuable information to unravel the intricancies of
the pion
nucleus interaction or the elementary properties of the $\Lambda
N$
interaction, as well as proper nuclear structure details of the
$\Lambda$
hypernuclei themselves.

\section{Weak non-mesonic decay of \L hypernuclei}

The pionic decay mode of \L hypernuclei is negligibly small in
medium to
heavy hypernuclei due to Pauli blocking acting on the final
nucleon which
emerges with a very small momentum. However, the medium effects
are
responsible for the opening of new decay channels,
where there are no pions in the final state.
These non-mesonic decay modes of $\Lambda$ hypernuclei can be
viewed as coming
from the mesonic decay when the pion is produced in a virtual
state
and absorbed by one or more nucleons. The one-nucleon induced
decay
mode corresponds to the \lnnn transition and can be interpreted
as being mediated by the exchange of one pion (or more massive
mesons).
In the two-nucleon induced decay of nuclei, the transition
$\Lambda N N \to NNN$ can be interpreted as  coming
from the absorption of the virtual pion
by a pair of nucleons that are
correlated by the strong interaction.
No matter what the interpretation of the different non-mesonic
decay
mechanisms is, what happens is that
the mass excess in the initial state (176 MeV)
is now converted
exclusively into kinetic energy of the final nucleons, which
may emerge with
large momentum values not restricted by Pauli blocking.
Hence, except for the very lightest ones, $\Lambda$ hypernuclei
decay
mainly through the non-mesonic mechanisms.

\subsection{One-nucleon induced decay}

The decay of hypernuclei through the one-body induced mechanism
\lnnn
offers the best opportunity to study the $\Delta S=-1$
nonleptonic
weak interaction between hadrons due to the practical
impossibility of having stable $\Lambda$ beams.
Since the typical momentum exchanged is around 400 MeV/c,
the process is short ranged and will not be too sensitive to the
nuclear structure details.
The \lnnn reaction, similar to the $\Delta S=0$ $NN\to NN$ weak
reaction but in the $\Delta S=-1$ sector, allows one to study not
only
the parity violating (PV) part of the interaction, but also the
parity conserving (PC) one which, in the
$NN\to NN$ case, is masked by the strong interaction.
Note, however, that the study of the inverse reaction $p n \to p
\Lambda$ with moderate energy and high intensity proton beams
could be feasible at RCNP (Osaka) \cite{kishi0} or COSY (J\"ulich)
\cite{haiden}.

The non-mesonic decay of hypernuclei has been studied for more
than thirty
years.
The experimental and theoretical status on the one-nucleon
induced
hypernuclear decay was extensively reviewed by Cohen in
1990 \cite{cohen1}.
For this reason we will just update the present situation in the
next paragraph and only the new achievements made since 1990 and
how they
compare with previous works will be covered in more detail in
the
coming sections.

Block and Dalitz \cite{dalitz} built a phenomenological model to
describe
the decay of s-shell hypernuclei in terms of a few elementary
spin-isospin reaction rates.
Although no assumptions were made on the dynamics of the \lnnn
reaction,
this model has been (and is) extremely useful because, by fitting
to
the empirical data for light hypernuclei, one can extract
information on the characteristics of the reaction,
such as the validity of the $\Delta I=1/2$
rule.
The first microscopic calculation was that of Adams \cite{adams}
who
considered the one pion exchange mechanism to describe the
non-mesonic
decay of a $\Lambda$ in nuclear matter and found a large
sensitivity
to the short range correlations (SRC) induced by the strong
interaction in the initial $\Lambda N$ and final $NN$ states.
A unified treatment of the mesonic and non-mesonic channels based
on Feynman diagrams and propagators was developed by Oset and
Salcedo \cite{oset}, where the medium effects in the
renormalization
of the pion were considered.
Being a short range process, it was clear that the one pion
exchange (OPE) mechanism might be
insufficient to explain the data.
McKellar and Gibson \cite{mckellar} made the first attempt to
include
the heavier mesons by developing a model for the exchange of the
$\rho$
meson. This was applied later to the decay of
light hypernuclei \cite{takeuchi} but the results depended on the
relative
sign between the $\pi$ and $\rho$ mechanisms that was not fixed
by the model.
Based on a pole model for the PC vertices and on
soft meson theorem techniques for the PV ones,
Dubach et al. \cite{dubach} constructed a full
one meson exchange
potential and preliminary results for nuclear matter were
obtained. The details of the model and final results
have recently become
available \cite{holstein}.  An extension of the model including
form
factors and some exchange diagrams related to the
antisymmetrization
of the
final two-nucleon state has been recently applied to the decay
of finite hypernuclei \cite{assum97}.
The exchange of $\sigma$ and $\rho$ mesons has also been
considered
within the image of correlated two pion exchange, where $\Delta$
and
$\Sigma$ are excited in the intermediate
state \cite{shmat94a,shmat94b,itonaga,iton97}.
A description of the \lnnn decay in terms of quark degrees of
freedom
was first attempted in refs.  \cite{kisslinger}, where a
hybrid
model that combined a quark Hamiltonian at short distances and
OPE at
long distances was employed. Oka and collaborators \cite{oka}
combined a direct quark mechanism with a one pion exchange
contribution for the
description of light hypernuclei. A more realistic $\Lambda$ wave
function
was used in a more recent work \cite{oka2}, where also the phase
between
the direct quark and the pion mechanisms was established.
One kaon exchange amplitudes, which partially cancelled the
pion ones, were also calculated in ref.  \cite{maltman} together
with
the quark contributions.
All the models and results of the recent works will be
extensively
reviewed below.

\subsubsection{OPE results}

\begin{figure}[htb]
       \setlength{\unitlength}{1mm}
       \begin{picture}(80,60)
      \put(18,-42){\epsfxsize=11cm \epsfbox{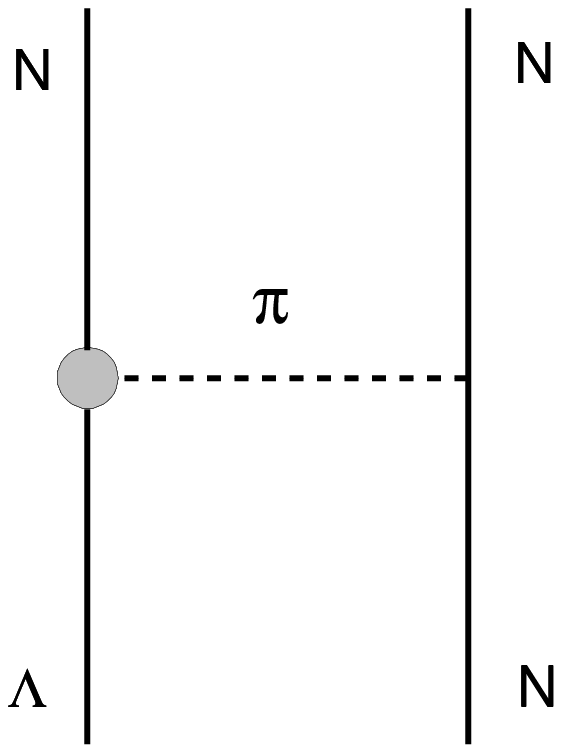}}
        \end{picture}
\caption{
One pion exchange diagram for the \lnnn transition
}
\label{fig:lnnn}
\end{figure}

The OPE mechanism for the \lnnn transition displayed
in
fig. \ref{fig:lnnn} involves a weak and a strong vertex.

As already shown in eq. (1), the weak Hamiltonian is
parametrized in the form
\begin{equation}
{\cal H}^{\rm W }_{\rm {\scriptscriptstyle \Lambda N} \pi}= i G_F
\mu^2
\overline{\psi}_{\rm N}
(A_\pi+B_\pi \gamma_5)
\vec{\tau} \vec{\phi}_\pi
\psi_\Lambda \, \left( ^0_1 \right) \ ,
\end{equation}
which implements the $\Delta I=1/2$ rule through the
isospin spurion $\left( ^0_1 \right)$.
The consequences of breaking this rule in the non-mesonic decay
of hypernuclei have been
recently
explored and will be presented in Sect. 3.1.4.

For the strong vertex the pseudoscalar coupling
\begin{equation}
{\cal H}^{\rm S }_{\rm {\scriptscriptstyle NN} \pi}= i g_{\rm
 {\scriptscriptstyle NN} \pi}
\overline{\psi}_{\rm N}
\gamma_5
\vec{\tau}\vec{\phi}_\pi \psi_{\rm N} \ ,
\label{eq:strongpi}
\end{equation}
has usually been taken since most calculations are
nonrelativistic and
this vertex is equivalent to the pseudovector one. The
nonrelativistic transition
potential reads
\begin{equation}
V_{\pi}(\vec{q}\,) = - G_F \mu^2
\frac{g_{\rm {\scriptscriptstyle NN} \pi}}{2M} \left(
A_\pi + \frac{B_\pi}{2\overline{M}}
\vec{\sigma}_1 \vec{q}\, \right)
\frac{\vec{\sigma}_2 \vec{q}\, }{\vec{q}\,^2+\mu^2}
\vec{\tau}_1
\vec{\tau}_2
\label{eq:pion}
\end{equation}
where $\vec{q}$ is the momentum carried by the pion directed
towards the
strong vertex, $\mu$ the pion mass,
$M$ the nucleon mass and $\overline M$
the average between the nucleon and $\Lambda$ masses.

By introducing the tensor operator $S_{12}(\hat{r}) = 3
(\vec{\sigma}_1 \hat{q}\, )
(\vec{\sigma}_2 \hat{q}\, ) - \vec{\sigma}_1 \vec{\sigma}_2 $,
the OPE transition potential can be splitted into central,
tensor
(adding to a total PC part) and PV pieces.
The decay rate can be obtained in terms of $< f \mid V_\pi \mid i
>$,
the expectation value of the potential between initial $\Lambda
N$ and
final $NN$ two-body states.

There are a few relativistic corrections which can be
implemented. The vertices can be made relativistic (see eq. (32)
of \cite{ramos94} and eq. (3) of \cite{pedro95}) and the pion
propagator non-static, i.e., $\vec{q}\,^2 \to -q^2$. Furthermore,
one should use, for consistency, relativistic wave functions
\cite{ramos91,ramos92}. Both types of corrections are of the
order of 10\% and of opposite sign, so we shall ignore the
relativistic effects in what follows.

\vskip 0.2cm

{\bf a) Nuclear matter}
\addcontentsline{toc}{subsubsection}{\protect\numberline{~~~{
a)}}{Nuclear matter}}

\vskip 0.2cm
The first work to consider the OPE mechanism was that
of
Adams \cite{adams}, where the importance of strong and final
state
interactions in the \lnnn transition was first pointed
out. The calculation was made in nuclear matter and only the
$L=0$
$\Lambda N$ relative motion was retained.  The same approach was
followed by McKellar and Gibson \cite{mckellar} who made the
first
attempt
to improve on the OPE mechanism by including the
$\rho$ meson as well. This will be discussed in the next section.
Their uncorrelated one-pion rate was consistent with that of
Adams, after correcting the too small coupling constant, but
the effect of correlations was found less strong. This was
confirmed
by the two other nuclear matter calculations that appeared
at that time. Oset and Salcedo \cite{oset} used a Green's
functions formalism to treat the mesonic and non-mesonic channels
in a unified way. Their method includes automatically all the
partial waves of the relative $\Lambda N$ motion.
The OPE mechanism was also studied by Dubach and
collaborators
in the context of a one meson exchange model for the \lnnn
transition \cite{dubach,holstein}.

The results for the one-nucleon induced non-mesonic decay rate in
nuclear
matter
are summarized in Table \ref{tab:nucmat}. The values of this
table, and all others referring to decay rates, will be given
in units of the free $\Lambda$ width, $\Gamma_\Lambda = 3.80
\times 10^9$ s$^{-1}$. The uncorrelated
results obtained
by the various groups are very similar. Substantial
diferences are found when SRC
are considered, although Adams used an unusually strong tensor
correlation that, when omitted, would bring the correlated result
up to 1.57.
The discrepancy of a factor of 2 between refs.  \cite{oset}
and \cite{mckellar},
when form factors and short range effects tied to the initial
and final strong correlations are included, can
be understood from differences in the model ingredients.
In ref.  \cite{mckellar} a combined form factor of the type
$\Phi(\vec{q}\,)=(\Lambda_\pi^2-\mu^2)/(\Lambda_\pi^2+\vec{q}\,^2
)$
is
employed
with $\Lambda_\pi^2=20\mu^2$, while ref.  \cite{oset} uses this
monopole form factor at each vertex with $\Lambda_\pi=1300$ MeV.
For the relevant momentum transfer of 400 MeV/c, the ratio
of form factors squared is 1.8, while the
different correlation functions account for the
remaining difference.
McKellar and Gibson use a Gaussian type,
$1-{\rm exp}(-r^2/b^2)$, with $r=0.75$, while Oset and Salcedo
use a Bessel type, $1-j_0(q_c r)$, with $q_c=3.93$ fm$^{-1}$.
As seen in ref.  \cite{assum1}, the rates calculated with
these two choices differ by about 30\% and the Bessel type
is closer to what is obtained from microscopic G-matrix
calculations
using the Nijmegen $\Lambda N$ interactions.

\begin{table}[htb]
\centering
\caption{ One pion exchange contribution to the
$\Lambda$ non-mesonic
decay rate in nuclear matter.
}

\begin{tabular}{|l|c|c|c|}
\hline
Group & uncorr. & corr. (no FF) & corr. + FF \cr
\hline
Adams\protect \cite{adams} & 3.47$^*$ & 0.38 &  \cr
McKellar-Gibson\protect \cite{mckellar} & 4.13 & 2.31 & 1.06 \cr
Oset-Salcedo\protect \cite{oset} & 4.3 & & 2.2 \cr
Dubach et al.\protect \cite{holstein} & 4.66 & 1.85 &  \cr
\hline
\end{tabular}

$^*$ {\small Corrected by a factor of 6.81 as pointed out in
ref.  \cite{mckellar}}

\label{tab:nucmat}
\end{table}

In summary, the uncorrelated OPE contribution to the $\Lambda$
decay rate in nuclear
matter turns out to be around 4 times the free $\Lambda$ width
and
the results are lowered by a factor of 2 when appropriate
short range correlations in the
initial and final state are considered.

\vskip 0.2cm
{\bf b) Finite nuclei}
\addcontentsline{toc}{subsubsection}{\protect\numberline{~~~{
b)}}{Finite nuclei}}

\vskip 0.2cm

The existence of new experimental data for the non-mesonic
decay rates \cite{szymanski,noumi} and the possibility of
obtaining
polarization observables, such as the asymmetry in the
distribution
of protons emerging from the weak decay of \hyp{12}{C}
\cite{ajim}
motivated
a new calculation of the non-mesonic decay performed directly for
the finite hypernuclear system \cite{ramos91,ramos92}. The
formalism
was chosen relativistic in view of the success of this model in
reproducing
polarization observables in proton nucleus scattering reactions.
The \lnnn transition was obtained from the relativistic Feynman
amplitude
\begin{equation}
t_{fi} = i G_F \mu^2
g_{\rm {\scriptscriptstyle NN}\pi}
\int d^4x d^4y
\overline{\Psi}_{\vec{k}_1}(x) (A + B \gamma_5)
\Psi_{\alpha_\Lambda}(x)
\Delta_\pi(x-y) f(\mid \vec{x} - \vec{y} \mid)
\overline{\Psi}_{\vec{k}_2}(y) \gamma_5
\Psi_{\alpha_{\rm N}}(x) \ ,
\label{eq:relati}
\end{equation}
where the initial \L and nucleon were solutions of the Dirac
equation
with scalar and vector potentials, chosen to reproduce the
nucleus
form
factor and the nucleon and \L separation energies. The two
emerging
nucleons felt the effect of a relativistic optical potential.
The short range effects were considered via a
phenomenological
correlation function, $f$, modifying the pion propagator
$\Delta_\pi$.
The value of the free rate for
\hyp{12}{C} was found to be about a factor of
2 smaller
than the nuclear matter results. This is reasonable because a
finite
hypernucleus has low density regions which give a smaller
contribution to the
decay rate. It was surprising, however, that the rate was reduced
by a factor of four
due to SRC and FF at the vertices, an effect twice as large as
that
observed in the nuclear matter calculations.
To adscribe the differences to the finite size of
the system is not reasonable because correlations are short
ranged
and they should affect similarly the infinite and finite systems.
Another possibility would be to relate the discrepacies to the
relativistic model but ref.  \cite{ramos91} already showed that,
in the absence of correlations, the relativistic and
nonrelativistic
treatments were giving the same results to within 10\%.

\begin{figure}[htb]
       \setlength{\unitlength}{1mm}
       \begin{picture}(100,60)
      \put(30,-40){\epsfxsize=9cm \epsfbox{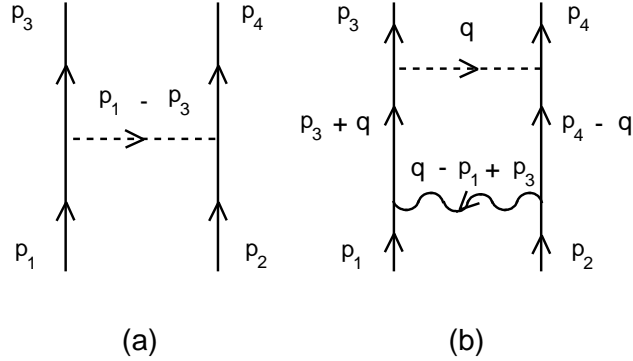}}
       \end{picture}
\caption{
Schematic model for correlations in the OPE potential.
Bare OPE diagram (a) and OPE with simultaneous
exchange
of a heavy meson (b).
}
\label{fig:correl}
\end{figure}

The answer to this problem was partially given in ref.
 \cite{sitges},
where it was shown, through a nonrelativistic reduction of the
correlated Feynman amplitude of eq. (\ref{eq:relati}),
that one did not obtain the same correlated potential as the one
used in the nonrelativistic approaches, namely $f(r)
V_{\pi}(\vec{r})$.
The final answer was given in ref.  \cite{assum1} where, through
a
microscopic model for correlations, it was shown
that the correct procedure is that followed by the
nonrelativistic
models.
The essential point is that the origin of correlations is tied
to the exchange of a heavy meson, as e.g. the $\omega$ meson,
represented
by the wavy line in fig. \ref{fig:correl}. Then the vertices
involved in
the weak transition mediated by the pion (dashed line)
connect
final state spinors with intermediate spinors which depend on the
momentum over which one must integrate. This can be clearly seen
from
the momentum space expression corresponding to the PC piece of
the
weak transition potential
\begin{equation}
\int \frac{d^3q}{(2\pi)^3} D(\vec{q}\,) \bar{u}(\vec{p}_3)
\gamma_5
u(\vec{p}_3+\vec{q}\,) \bar{u} (\vec{p}_4) \gamma_5 u(\vec{p}_4 -
\vec{q}\, ) \dots
\end{equation}
where $D(\vec{q}\,)$ is the pion propagator and the dots
simbolize
additional elements of the amplitude which are not relevant for
the
discussion. However, the relativistic procedure followed in
refs.  \cite{ramos91,ramos92} was equivalent to working with
external momentum spinors which, therefore, factorize out of the
integral
\begin{equation}
 \bar{u}(\vec{p}_3) \gamma_5
u(\vec{p}_1) \bar{u} (\vec{p}_4) \gamma_5 u(\vec{p}_2)
\int \frac{d^3q}{(2\pi)^3} D(\vec{q}\,)\dots
\end{equation}
In coordinate space, this procedure leads to oscillating
potentials \cite{assum1} which produce an artefactual reduction
of the rates.

The calculated non-mesonic decay rates of several hypernuclei have been
reported recently \cite{assum97} in
the framework of a nuclear shell-model and a nonrelativistic
transition operator. Details on how the hypernuclear
transition amplitude can be decomposed in terms of two-body
amplitudes for nucleons in any shell can be found in
\cite{tesi97}.
The nucleons and the \L are described with harmonic oscillator
wave
functions, SRC are included via a
phenomenological correlation function inspired in
G-matrix calculations with realistic $\Lambda N$
interactions and final state interactions (FSI) are treated
in terms of two-body $NN$ scattering states obtained
from realistic $NN$ interactions.
A phenomenological form factor is included at each vertex as
well.
The effect of the different ingredients is visualized in
Table \ref{tab:opefin1} for the decay rate of \hyp{12}{C},
splitted into central, tensor (adding to a total PC rate) and PV
contributions.

\begin{table}[hbt]
\begin{center}
\centering
\caption{ OPE contribution to the non-mesonic decay rate
of \hyp{12}{C} from ref. \protect\cite{assum97}  }
\begin{tabular}{|l|c|c|c|c|c|}
\hline
   & FREE & SRC & SRC+FF & \multicolumn{2}{c|}{SRC+FF+FSI} \\
   &   &   &  & phenom. & Nijm93  \\
\hline
Central rate  & 0.282 & $3.4 \times 10^{-3}$ & $1.3 \times 10^{-2}$ &
$4.2 \times 10^{-3}$ & $3.3 \times 10^{-3}$ \\
Tensor rate  & 0.858 & 0.781 & 0.637 & 0.685 & 0.566  \\
PC & 1.140 & 0.785 & 0.650 & 0.689 & 0.531  \\
PV & 0.542 & 0.447 & 0.389 & 0.421 & 0.353  \\
\hline
$\Gamma/\Gamma_\Lambda$ & 1.682 & 1.232 & 1.038 & 1.110 & 0.885
\\
\hline
\end{tabular}
\label{tab:opefin1}
\end{center}
\end{table}

The free central rate is reduced drastically by SRC, however,
most of its contribution comes from the delta function which
is completely eliminated by the correlation function,
which is zero at the origin. The tensor rate is reduced by
10\% by SRC and by 20--35\% once FF and FSI are included.
The PV rate represents  40\% of the total rate,
differing from previous nuclear matter results, which
reported either negligible \cite{mckellar} or  15\% \cite{dubach}
contribution to the rate. However, a simple estimate of the PV
contribution, relative to the PC one, for the most relevant
momentum transfer (400 MeV/c) gives
$$\left(\frac{A_\pi}{\frac{B_\pi {\rm q}}{2
\overline{M}}}\right)^2 = 0.57 \ ,$$
which yields a 35\% fraction of the total for the PV rate.
The last two columns reflect the
importance of using a realistic $NN$ wave function in the
final state. A phenomenological $NN$ correlation function
of the type $1-j_0(q_c r)$, with $q_c=3.93$ fm$^{-1}$, gives a
rate
25\% larger than that obtained from  a $NN$ wavefunction based on
the
Nijmegen93 potential \cite{stoks94}.
The $NN$ interaction is relevant because it modulates the wave
function at the short relative distances and prevents the
contribution from the short part of the interaction. One may
wonder about the interaction of the emitted nucleons with the
mean potential of the nucleus. This can be taken into account by
means of a complex optical potential, the real part of which is
moderate and has practically no effect in the transition rate.
The imaginary part would induce a loss of flux due, in the
present case, to $NN$ collisions. However, the net effect is a
redistribution of the original $NN$ emission strength into
different multinucleon channels \cite{Ca94} without modifying the
original decay rate, which is the magnitude in which we are
interested.

One of the interest in doing finite nuclei calculations
was to establish the role of relative $L \neq 0$ motion
in the non-mesonic weak decay. It was especulated that the
p-shell
rate would be basically coming from relative $\Lambda N$
motion in $P$ wave, due to supression of $S$-wave by the
corresponding $L=1$
center-of-mass wave function which is zero at the
most probable back-to-back kinematics.
However, it was found \cite{bennhold2} that, after integrating
over all
possible angular possibilities,
90\%  of the p-shell contribution comes from the relative
$\Lambda N$
$L=0$ motion, which has been confirmed by the most elaborated
model of ref.  \cite{assum97}.

In Table \ref{tab:opefin2} finite nuclei results from different
works are compared.
The results within brackets, if available, correspond to the
calculation omitting SRC, FF and FSI, which will be referred to
as the uncorrelated ones.

\begin{table}[hbt]
\centering
\caption{ OPE contribution to the non-mesonic decay rate of
hypernuclei. Results within brackets correspond to the
calculation omitting SRC, FF and FSI.}
\begin{tabular}{|l|c|c|}
\hline
 & \hyp{5}{He} & \hyp{12}{C} \cr
\hline
Takeuchi et al. \protect \cite{takeuchi} & (0.52) 0.14  &  \cr
Dubach et al. \protect \cite{holstein} & (0.6) 0.9   & (3.4) 0.5
\cr
 & (1.6) \phantom{0.9}  & \phantom{(3.4)~~~~~}
2.0 \protect \cite{dubachwein} \cr
Parre\~no et al. \protect \cite{assum97} & (0.98) 0.5  & (1.68)
0.89
\cr
Oset-Salcedo \protect \cite{oset} & \phantom{(0.98)} 1.15 &
\phantom{(1.68)}     1.45 \cr
\hline
EXP: & $0.41\pm 0.14$ \cite{szymanski} & $1.14 \pm
0.2$ \cite{szymanski}
\cr
    &                             & $0.89 \pm 0.18$ \cite{noumi}
\cr
\hline
\end{tabular}
\label{tab:opefin2}
\end{table}

Compared to the other predictions, a quite
low rate for \hyp{5}{He} is found in ref.  \cite{takeuchi}. Part
of it
is due to the use of a
more realistic \L wave function pushed to the surface by the
effect of
the
repulsive $\Lambda$ nucleus interaction in light hypernuclei.
But most of the reduced rate comes from a much stronger effect of
FF
and SRC as they use the model of ref.  \cite{mckellar} discussed
in the nuclear matter section. The recent results of Dubach
et al.  \cite{holstein} are somewhat confusing and the lack of
computational
details makes the comparison with other works difficult. However,
some
general points are worth mentioning. Their uncorrelated rate for
\hyp{12}{C} is a factor 2 larger than that of  \cite{assum97} and
the correlated one a factor 2 smaller, so the effect of
correlations
is substantially larger than that found in their nuclear matter
results. It is also surprising that their correlated rate for
\hyp{5}{He}
is larger than the uncorrelated one and a factor of 2 larger than
the \hyp{12}{C} result. An explanation could lie in the
possibility of some misprints. Their \hyp{12}{C} rate quoted more
than
10 years ago in some conference proceedings \cite{dubachwein}
turns out to
be
around 2. Moreover, the preprint of their most recent
paper \cite{holstein}
quotes 1.6 for the uncorrelated \hyp{5}{He} result. This two
corrections
would lead to more consistent results and would differ only
in a factor of two from those of ref.  \cite{assum97}. We note
that the model of ref.  \cite{holstein} does not contain
form factors at the vertices, which would provide some
additional reduction.
The LDA results of ref.  \cite{oset}
appear to be a little bit too high when compared
to
the direct calculations on finite nuclei. However, we have
checked
that, when using the same
\L wave function as in ref.  \cite{assum97} (slightly more
extended),
a similar correlation function ---controlled in  \cite{oset}
by the Landau $g^\prime_\Lambda$ parameter---,
and neglecting the pion renormalization, the LDA result
lowers down to 1.05, nicely agreeing with the
value 1.038 quoted in Table \ref{tab:opefin1} without FSI
effects,
which are absent in  \cite{oset}.
This consistency check reinforces the idea that the LDA
is an excellent tool to study processes that are not too
sensitive to nuclear structure details or low energy
nuclear excitations as is the case of the non-mesonic
weak decay of hypernuclei.

After this comparison, we can safely claim that, in general, the
OPE
mechanism
predicts non-mesonic decay rates of finite nuclei in agreement
with the experimental values.

\begin{table}[htb]
\centering
\caption{ OPE results for the ratio $\Gamma_n/\Gamma_p$}
\begin{tabular}{|l|c|c|}
\hline
 & $\pi$ (uncorr.) & $\pi$ (corr.)   \cr
\hline
Takeuchi et al.  \cite{takeuchi} (\hyp{5}{He})& & 0.06  \cr
Inoue et al.  \cite{oka} (\hyp{5}{He})&  & 0.12 \cr
\phantom{Inoue et al. \cite{oka}} (\hyp{4}{He}) &   & 0.08 \cr
Parre\~no et al.  \cite{assum97} (\hyp{12}{C})& 0.18 &  0.10 \cr
\phantom{Parre\~no et al.  \cite{assum97}} (\hyp{5}{He})& 0.12 &
0.07 \cr
Dubach et al.  \cite{holstein} (nuc. matt.) & 0.09 & 0.06  \cr
\phantom{Dubach et al.  \cite{holstein}} (\hyp{12}{C}) & 0.22 &
0.20  \cr
\phantom{Dubach et al.  \cite{holstein}} (\hyp{5}{He)} & 0.07 &
0.05  \cr
\hline
\multicolumn{2}{|r|}{EXP
\phantom{AAAAAAAAAAAA}\hyp{4}{He}\phantom{AAA}} & $0.06 \pm 0.30$
\cite{outa97}
\cr
\multicolumn{2}{|l|}{~} & $0.25 \pm 0.13$ \cite{zeps97} \cr
\multicolumn{2}{|r|}{\hyp{5}{He}\phantom{AAA}} & $0.93 \pm 0.5$
\cite{szymanski}
\cr
\multicolumn{2}{|r|}{\hyp{12}{C}\phantom{AAA}} &
$1.33^{+1.12}_{-0.81}$ \cite{szymanski}
\cr
\multicolumn{2}{|l|}{~}  & $1.87^{+0.91}_{-1.59}$ \cite{noumi}
\cr
\multicolumn{2}{|l|}{~}
  & $0.70 \pm 0.30$ \cite{montwill} \cr
\multicolumn{2}{|l|}{~}
  & $0.52 \pm 0.16$ \cite{montwill} \cr
\hline
\end{tabular}
\label{tab:opefin3}
\end{table}

However, the OPE model is unable to explain
the ratio $\Gamma_n/\Gamma_p$ between the neutron-induced rate
($\Lambda n \to nn$)
and the proton-induced one ($\Lambda p \to n p$).
In Table \ref{tab:opefin3} several predictions for this ratio
are compared with the experimental
results.
While the
data, with large error bars, seem to suggest a value of around
1 or larger (except for the very light hypernucleus
\hyp{4}{He}  \cite{outa97,zeps97}), all models give values around
0.05--0.2. This is a well known
characteristic of the
OPE mechanism, which favors the tensor transitions over the
central and parity violating ones. Since the $\Lambda N$ pair
is mainly moving in $L=0$ states and the tensor operator
induces ${}^3S_1 (\Lambda N) \to {}^3 D_1 (NN)$ transitions,
the antisymmetric final system will be predominantly in an
isospin I=0
state, which only $np$ pairs can achieve. Hence, the
transition $\Lambda n \to n n $ is highly suppressed
in the OPE model. These arguments have been more quantitatively
stated in ref. \cite{oset90}, within the propagator method. In
the limit of zero momentum for the $\Lambda$ and small Fermi
momentum for the nucleons, the ratio $\Gamma_n/\Gamma_p$ is 1/14
when the antisymmetry of the final two nucleons is taken into
account. The ratio becomes 1/5 when the antisymmetry is
neglected, while the total width is changed only at the level of
20\%.

This discrepancy motivated the search for new mechanisms
driving the $\Lambda N \to N N$ decay, such as the inclusion
of other mesons, correlated two pion exchange, quark model
calculations, incorporation of $\Delta I=3/2$ amplitudes, etc,
all of which will be described in coming sections.

\subsubsection{Meson exchange model beyond OPE}

The OPE mechanism is an excellent starting point to study the
weak \lnnn decay due to the experimentally known
weak vertex. However, this mechanism can at most describe the
long range part of the transition potential, while it was soon
realized \cite{adams} that the large momentum exchanged (400
MeV/c)
would probe quite short distances and more massive mesons might
contribute in the decay process.

\vskip 0.2cm
{\bf a) The $\rho$ meson}
\addcontentsline{toc}{subsubsection}{\protect\numberline{~~~{
a)}}{The $\rho$ meson}}
\vskip 0.2cm

Including the $\rho$ meson in the weak decay mechanism was fist
attempted by McKellar
and Gibson \cite{mckellar}, where the weak vertex Hamiltonian
was
\begin{equation}
{\cal H}^{\rm W}_{\rm {\scriptscriptstyle \Lambda N} \rho} = G_F
\mu^2 \:
{\overline \psi}_{\rm N} \: \left( \alpha \gamma^\mu
- \beta i \frac{\sigma^{\mu \nu} q_\nu} {2 \overline{M}} +
\varepsilon
\gamma^\mu \gamma_5 \right)
\vec{\tau}\vec{\rho}_\mu \psi_\Lambda \, \left( ^0_1
\right) \ ,
\end{equation}
where, contrary to the pion, the weak constants $\alpha$, $\beta$
and $\varepsilon$ must be determined theoretically. Taking the
usual
strong $\Lambda N \rho$ vertex
\begin{equation}
{\cal H}^{\rm S}_{\rm {\scriptscriptstyle NN} \rho} =
\overline{\psi}_{\rm N}
\left( F_1
 \gamma^\mu + i
\frac{ F_2}{2M}
\sigma^{\mu \nu} q_\nu \right)
\vec{\tau}\vec{\rho}_\mu \psi_{\rm N}   \ ,
\end{equation}
the nonrelativistic reduction
of the $\rho$ transition potential reads
\begin{eqnarray}
{V_{\rho}}(\vec{q}\,)  &=&
G_F \mu^2
 \left( F_1  \alpha - \frac{(\alpha + \beta )
 ( F_1 + F_2 )} {4M \overline{M}}
( \vec{\sigma}_1 \times \vec{q}\,)
(\vec{\sigma}_2 \times \vec{q}\,) \right. \nonumber \\
& & \phantom { G_F m_\pi^2 A }
\left. +i \frac{\varepsilon ( F_1 + F_2 )} {2M}
(\vec{\sigma}_1 \times
\vec{\sigma}_2 ) \vec{q}\,\right)
\frac{1}{\vec{q}\,^2 + m_\rho^2} \ .
\label{eq:rhopot}
\end{eqnarray}
Using
$(\vec{\sigma}_1 \times \vec{q}\,) (\vec{\sigma}_2
\times \vec{q}\,) =
(\vec{\sigma}_1 \vec{\sigma}_2)
\: \vec{q}\,^2 - (\vec{\sigma}_1 \vec{q}\,)
 (\vec{\sigma}_2 \vec{q}\,)$, and decomposing
$(\vec{\sigma}_1 \vec{q}\,) (\vec{\sigma}_2
\vec{q}\,)$ into central and tensor operators, this potential
contains, as in the case of pion exchange, central, tensor and
parity violating
pieces.

Two weak coupling models are used in \refer{mckellar}, namely, a
modified factorization approximation ---omitting the factor
$\sin{\theta_c} \cos{\theta_c}$--- and a pole model which applies
only to the PC amplitudes. The resulting coupling
constants differ substantially in both approaches, the pole model
giving a combination $(\alpha + \beta)$ which differs in sign and
is a factor of 20 smaller than that obtained in the factorization
approximation. Combining the OPE results with the $\rho$
contribution
with an arbitrary relative sign, gives, for the factorization
approach, a non-mesonic rate which ranges between 0.7 (for $\pi -
\rho$) and 2.3 (for $\pi + \rho$). Aware of the limitations of
the
factorization model, the authors of \refer{mckellar} state that
their estimates can vary up to a factor 3. An important
observation was that, similarly as for their OPE results, only
the tensor transition gave a non-negligible contribution to the
$\rho$ exchange weak amplitude.

Despite the large uncertainty in the result, an application of
this model to finite nuclei was performed in \refer{takeuchi},
where the decay rate of helium hypernuclei was explored. The
$(\pi + \rho)$ model gave an amplification of their $\pi$-only
result and the ($\pi - \rho$) one lead to a complete
cancellation. Although the \hyp{5}{He} result was in nice
agreement with the data in the $(\pi + \rho)$ model, the
\hyp{4}{He} rate was overestimated by a factor of 2.

Nardulli \cite{nardulli} obtains the weak $\rho$ couplings via a
pole model, including, in addition to the ground state baryon
octet,
the first excited $1/2^+$ baryons. However, it must be
recalled \cite{maltman95a}
that their value for the $K \to \pi$ transition is an order of
magnitude larger than what is obtained from the kaon decay mode
$K \to \pi \pi$ \cite{donoghue}. The PV couplings were also
calculated within a pole model including the negative parity
$1/2^-$ baryon poles. By rescaling the results of
\refer{mckellar} according to the new coupling constants,
Nardulli predicts the non-mesonic decay rate of a \L in nuclear
matter to be 0.7.

The $\rho$ meson was also investigated in \refer{holstein} as a
part of their full one meson exchange potential, which will be
discussed in more detail below. Only the ground state baryons are
considered in their determination of the PC coupling constants,
while the PV ones are obtained applying soft meson theorems and
current algebra techniques. Although the baryon pole diagrams
could, in principle, also contribute to the PV couplings, they
were estimated to be only a several percent of the leading
current algebra contribution \cite{donoghue}. The contribution of
the $\rho$ meson, acting mainly on the $S\to D$ transition, was
found to decrease the pion-only rate by 15\%.

A thorough study of the $\rho$-meson contribution to the decay
rate
of \hyp{12}{C} was performed recently \cite{assum95}, using the
model of \refer{holstein} for the weak $\Lambda N \rho$ couplings
and employing strong couplings from realistic $Y N$
potentials \cite{nijmegen,juelich}.
One of the findings of that work is that the central amplitude
of the $\rho$-meson contribution cannot be neglected and is, in
fact, larger than the tensor one. The reason is
that the function
$j_2(kr)$, representing the
outgoing relative $NN$ wavefuntion,
eliminates strength of the tensor potential from the short
distances. Due to its shorter range, the tensor contribution
of the $\rho$ is strongly reduced with respect to the central
transitions, which are governed by the function $j_0(kr)$.
As can be seen in Table \ref{tab:rho} the central rate for the
$\rho$ contribution turns out to
be 3.5 times the tensor one, once one includes SRC, FF and FSI
(considered in a simplified way through the $NN$ correlation
function $f(r)=1-j_0(q_c r)$, with $q_c=3.93$ fm$^{-1}$). The
table shows that the $\rho$-meson reduces the
$\pi$-only rate by 10\%, similar to the 15\% reduction reported
in \refer{dubach}, although a large fraction of the
$\rho$-rate came from $S\to D$ transitions in the latter case.
Similar qualitative
conclusions are obtained when more realistic $NN$ wavefunctions,
based on a $T$-matrix calculation employing the Nijmegen93
potential \cite{stoks94}, are used \cite{assum97}. There, the
central to
tensor relative strength of the $\rho$ meson is 1.5 and it
reduces the $\pi$-only rate by 3\%.

\begin{table}[hbt]
\centering
\caption{ The $\rho$, $\pi$ and $\pi+\rho$ contributions to the
decay rate of \hyp{12}{C} from ref. \protect\cite{assum95}.
}
\begin{tabular}{|l|cc|c|c|}
\hline
 & \multicolumn{2}{c|}{$\rho$} & $\pi$ & $\pi + \rho$ \cr
 & uncorr. &  corr. & &   \cr
\hline
Central & 0.658 & 0.085 & 0.004 & 0.059 \cr
Tensor  & 0.155 & 0.024 & 0.685 & 0.454 \cr
PV      & 0.109 & 0.011 & 0.421 & 0.477 \cr
Total   & 0.921 & 0.119 & 1.110 & 0.991 \cr
\hline
\end{tabular}
\label{tab:rho}
\end{table}

It is clear that other conclusions could be reached had one used
a different model for the $\Lambda N \rho$ vertex. However, since
the central and tensor potentials are weighted by the same
$(\alpha + \beta)$ combination, it is quite likely that both
central and tensor amplitudes are important and need to be
included before any further improvement on the potential is made.

\vskip 0.2cm
{\bf b) The full one meson exchange potential}
\addcontentsline{toc}{subsubsection}{\protect\numberline{~~~{
b)}}{The full one meson exchange potential}}
\vskip 0.2cm

\begin{figure}[bht]
       \setlength{\unitlength}{1mm}
       \begin{picture}(100,60)
      \put(25,-30){\epsfxsize=10cm \epsfbox{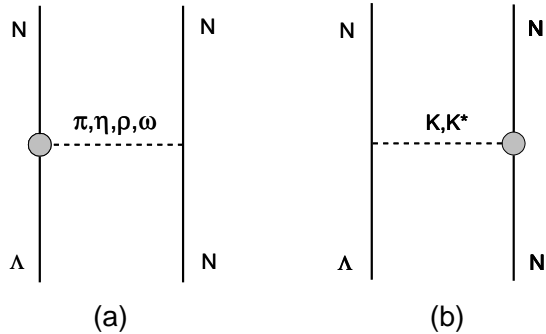}}
        \end{picture}
\caption{
Non-strange (a) and strange (b) meson exchange contribution to
the
$\Lambda N \to N N$ weak transition potential.
The weak vertex is indicated by the circle.
}
\label{fig:mesons}
\end{figure}

\begin{figure}[htb]
       \setlength{\unitlength}{1mm}
       \begin{picture}(100,100)
      \put(30,-3){\epsfxsize=9cm \epsfbox{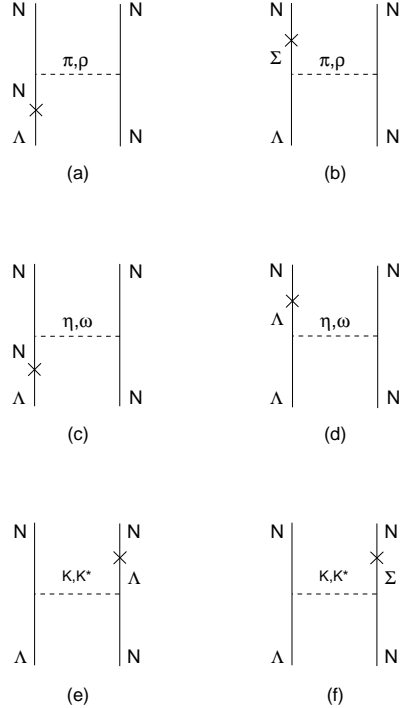}}
       \end{picture}
\caption{
Baryon pole diagrams contributing to the PC weak vertices in the
$\Lambda N \to N N$ transition amplitude.
}
\label{fig:pole1}
\end{figure}

\begin{figure}[hbt]
       \setlength{\unitlength}{1mm}
       \begin{picture}(80,60)
      \put(30,-25){\epsfxsize=9cm \epsfbox{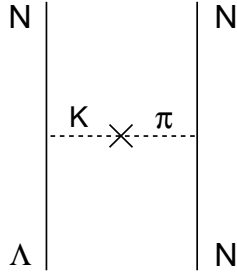}}
       \end{picture}
\caption{
Meson pole diagram contributing to the $\Lambda N \to N N$
transition amplitude.
}
\label{fig:pole2}
\end{figure}

A one meson exchange model to describe the \lnnn transition was
first reported in \refer{dubach} more than 10 years ago, and the
details have recently become available \cite{holstein}. As shown
in fig. \ref{fig:mesons},
pseudoscalar ($\pi,\eta,K$) and vector ($\rho,\omega,K^*$) meson
exchanges were included and the $\Delta I=1/2$ rule at
the weak vertex was assumed. The PC part of the vertices was
obtained
from a pole model including both baryon and meson pole diagrams
shown, respectively, in figs. \ref{fig:pole1} and
\ref{fig:pole2}, where the cross indicates a weak
baryon$\rightarrow$baryon or meson$\rightarrow$meson weak
transition. Employing SU(6)$_w$ symmetry and assuming PCAC,
current algebra methods allow one to relate the
baryon$\rightarrow$baryon amplitudes to the experimentally known
$\Lambda$ and
$\Sigma$ decay p-wave amplitudes, and the meson$\rightarrow$meson
ones to the $K \to \pi$ one, which is determined from the
physical $K \to \pi \pi$ decay rate.
The strong couplings for the pseudoscalar mesons are obtained
from SU(3) symmetry and the Goldberger-Treiman relation, while
those of the vector mesons require the vector dominance
assumption.

This model was applied to obtain the non-mesonic decay rate of a
$\Lambda$ in nuclear matter \cite{dubach,holstein} and, later,
predictions were also given for \hyp{5}{He} and
\hyp{12}{C} \cite{holstein}, although no details are yet
available.

Following the lines of refs.  \cite{dubach,holstein}, a full one
meson exchange potential has recently been constructed to study
the non-mesonic decay rate of finite nuclei \cite{assum97}, where
the nuclear structure details of the hypernucleus under
consideration are carefully taken into account. A convenient
compact expression for the potential is given by
\begin{equation}
V(\vec{r}) =\sum_{i} \sum_\alpha V_\alpha^{(i)}
(\vec{r}) = \sum_i \sum_{\alpha}
V_\alpha^{(i)} (r) \hat{O}_\alpha \hat{I}_\alpha^{(i)}
\end{equation}
where the index $i$ runs over the different mesons
$(\pi,\eta,K,\rho,\omega,K^*)$ and $\alpha$ over the spin
operators $\hat{O}_\alpha \in \{
\hat{1}$ (Central spin independent), $\vec{\sigma}_1
\vec{\sigma}_2$ (Central spin dependent), $S_{12}$ (Tensor),
$\vec{\sigma}_2 \hat{r}$ (PV for pseudoscalar mesons) and
$(\vec{\sigma}_1\times \vec{\sigma}_2 ) \hat{r}$ (PV for
vector mesons)$\}$.
The isospin operator $\hat{I}_\alpha^{(i)}$ is
$\hat{1}$ for isoscalar mesons ($\eta,\omega$),
$\vec{\tau}_1 \vec{\tau}_2$ for isovector mesons $(\pi,\rho)$ and
a combination of $\hat{1}$ and $\vec{\tau}_1 \vec{\tau}_2$
for isodoublet mesons $(K,K^*)$.

The detailed form of the potential and the explicit values of the
coupling constants can be found in refs. \cite{assum97,tesi97}.
To facilitate
the comparison, it is worth pointing out the differences with
respect to the model of \refer{holstein}. In the first place, the
strong coupling constants employed in \refer{assum97} are taken
from realistic $YN$ interactions, such as the
Nijmegen \cite{nijmegen} or the J\"ulich  \cite{juelich} ones.
Secondly, the potential of \refer{holstein} does not contain form
factors in spite of the large momentum transferred in the process
which makes the pion to be far off-shell. Thirdly, a stronger
correlation function is used in \refer{holstein}, which in part
can compensate for the lack of form factors. The work of
\refer{assum97} uses
a parametrization\cite{sitges} which lies in between realistic
correlation
functions obtained from $G$-matrix calculations employing the
Nijmegen soft- and hard-core potentials. Finally,
the contribution of the meson pole diagrams to the PC couplings
were found to be small and, hence, they were omitted in
\refer{assum97}.

Results obtained from both models are shown in Table
\ref{tab:mesons1}, were the $\Lambda$ decay rate in nuclear
matter of \refer{holstein} is compared to that in \hyp{12}{C} of
\refer{assum97} for a sequential (and arbitrary) addition of the
different mesons. Although the reduction with respect to the
pion-only rate is similar when all the mesons are included (25\%
in the case of \refer{holstein}
and 18\% in the case of \refer{assum97}), a stronger sensitivity
to the
strange mesons is found in \refer{assum97}.

\begin{table}[hbt]
\centering
\caption{ One meson exchange results for the non-mesonic decay
rate.
}
\begin{tabular}{|l|c|c|}
\hline
Meson & \refer{holstein} &  \refer{assum97}   \cr
    &   (nuc. matt.) & \hyp{12}{C} \\
\hline
$\phantom{+ }\pi$ & 1.85 & 1.11 \\
$+ \eta$ & 1.68 & 1.06 \\
$+ K$ & 1.31 & 0.54 \\
$+\rho$ & 1.11 & 0.51 \\
$+ \omega$ & 1.19 & 0.58 \\
$+ K^*$ & 1.38 & 0.91 \\
\hline
\end{tabular}
\label{tab:mesons1}
\end{table}

The work of \refer{assum97} includes the mesons
by isospin pairs [$(\pi,\rho)$, $(K,K^*)$, $(\eta,\omega)$] and
finds
important destructive interference effects for each isospin-like
pair in the tensor transitions, which provide the most
important contribution for pseudoscalar mesons. On the contrary,
the PV rates tend to interfere constructively. The study also
demonstrates the effect of including realistic $NN$ scattering
wave functions in the final state, reducing the pion-only rate of
Table \ref{tab:rho} from 1.11 to 0.89.
The final results of \refer{assum97} for the non-mesonic decay
rate of \hyp{12}{C} are shown in Table \ref{tab:mesons2}, using
the Nijmegen (J\"ulich) strong coupling constants. The ordering
of the mesons is different from the previous table and reflects
the isospin-pair study mentioned above.
Table \ref{tab:mesons2} shows the significance of SRC,FF and FSI
in the decay
rate, drastically reducing the influence of the heavier mesons
on the rate. The $K$ meson lowers the rate by almost 50\%,
reduction that is partly compensated by its isospin partner, the
$K^*$.
The final result is either 15\% smaller (using the Nijmegen
couplings) or 15\% larger (using the J\"ulich couplings) than the
pion-only result. This sensitivity to the strong couplings is
unfortunate since one expected to use the non-mesonic decay of
hypernuclei to learn about the weak vertex. More data on $YN$
scattering is highly desirable in order to reduce the
uncertainties on the strong coupling constants. The table also
shows the results when new $K$-meson weak couplings including
one-loop corrections to the leading order in Chiral Perturbation
Theory \cite{savage} are employed. Since the one-loop
corrections
decrease the tree-level p-wave amplitudes by 30\%, the influence
of
the $K$-meson is reduced substantially giving rise to an
increased rate. Note, however, that the effect of counterterms
was ignored in \refer{savage} and further investigations are
necessary. In any case, any theoretical improvement in the
determination of the weak and strong vertices can be easily
incorporated in the meson exchange model of \refer{assum97}.

\begin{table}[hbt]
\centering
\caption{ One meson exchange results 
for the non-mesonic rate
of \hyp{12}{C}, 
obtained in \protect\cite{assum97}
using the Nijmegen
(J\"ulich) strong couplings}
\begin{tabular}{|l|c|c|}
\hline
Meson & uncorr. &  corr.   \\
\hline
$\phantom{+ }\pi$ & 1.68 (1.68) & 0.89 (0.89) \\
$+ \rho$ & 2.06 (2.33) & 0.86 (0.83) \\
$+ K$ & 1.34 (1.70) & 0.50 (0.51) \\
$+ K^*$ & 2.84 (3.82) & 0.76 (0.90) \\
$+ \eta$ & 2.47 (3.82) & 0.68 (0.90) \\
$+ \omega$ & 2.30 (4.34) & 0.75 (1.02) \\
\hline
K-couplings & & 0.84 (1.10) \\
from $\chi PT$ & & \\
\hline
\end{tabular}
\label{tab:mesons2}
\end{table}

The available results for the non-mesonic weak decay of finite
nuclei in the one meson exchange model are displayed in Table
\ref{tab:mesons3}. The results for \hyp{12}{C} of
\refer{holstein} are extremely small when compared with their
previous nuclear matter results shown in Table \ref{tab:mesons1}.
They are a factor 7 smaller, which seems
unreasonable since one may expect reduction factors of at most
1.5 by going from nuclear matter to a finite nucleus as
\hyp{12}{C}.
However, the results of the same group quoted in
\refer{dubachwein},
shown between brackets in Table \ref{tab:mesons3}, seem to be
more consistent
with their previous nuclear matter results, with the calculations
of \refer{assum97} and with the experimental data.
Taking these considerations into account we can say that, in
general, the non-mesonic rates of finite
hypernuclei obtained within the
one meson exchange model are in fair agreement with the measured
values.

\begin{table}[hbt]
\centering
\caption{ Non-mesonic decay rates within the meson exchange
model of refs. \protect\cite{holstein} and
\protect\cite{assum97}.
}
\begin{tabular}{|l|cc|cc|}
\hline
 & \multicolumn{2}{c|}{\hyp{5}{He}} &
  \multicolumn{2}{c|}{\hyp{12}{C}} \\
 & $\pi$ & all mesons & $\pi$ & all mesons  \\
\hline
\refer{holstein}& 0.9 & 0.5 & 0.5 (2.0  \cite{dubachwein})&
0.2 (1.2  \cite{dubachwein}) \\
\refer{assum97} & 0.5 & 0.41 & 0.89 & 0.75 \\
\hline
EXP: & \multicolumn{2}{c|}{$0.41 \pm 0.14$ \cite{szymanski}} &
\multicolumn{2}{c|}{$1.14 \pm 0.2$ \cite{szymanski}} \\
 & & &
\multicolumn{2}{c|}{$0.89\pm 0.18$ \cite{noumi}} \\
\hline
\end{tabular}
\label{tab:mesons3}
\end{table}

The one meson exchange model has been recently applied to study
the non-mesonic decay of the hypertriton,
\hyp{3}{H} \cite{golak97}. The wave function of the hypertriton,
obtained by solving the Fadeev equations based on realistic $NN$
forces and the Nijmegen soft-core $YN$ interaction, was found
to reproduce nicely the $\Lambda$ separation energy,
$B_\Lambda=0.13\pm 0.05$ MeV
 \cite{miyagawa}.  The two non-mesonic decay modes, \hyp{3}{H}
$\to
d + n$ and \hyp{3}{H} $\to n n p$, were investigated and the
results are shown in Table \ref{tab:mesons4}. The one-pion only
result turns out to be 1.7\% of the free \L decay rate, which is
almost a factor 3 smaller than what was obtained in a simpler
calculation \cite{bennhold92}, which ignored FSI. Indeed, the
results in the table show that FSI have a strong influence on the
rate, reducing the PWIA value by a factor of 2.
Including all mesons decreases the one-pion result by around
10\%.
That work also makes extensive investigation of the kinematical
regions where protons and neutrons coming from the partial
neutron- and proton-induced mechanisms could be found. It was
shown
that the regions covered by the two mechanisms largely overlap
and, therefore, the partial rates cannot be measured separately
unless one applies kinematical detection constraints, which would
then lead to the measurement of only fractions of these partial
rates.

\begin{table}[hbt]
\centering
\caption{ Non-mesonic decay rate of \hyp{3}{H} from
\protect\cite{golak97}
}
\begin{tabular}{|l|c|c|}
\hline
 & $\pi$ & all mesons  \\
\hline
$\Gamma^{n + d}_{\rm PWIA}$& 0.0047 & 0.0037 \\
$\Gamma^{n + d}$& 0.0012 & 0.0017 \\
\hline
$\Gamma^{n + n + p}_{\rm PWIA}$& 0.036 & 0.028 \\
$\Gamma^{n + n + p}$& 0.018 & 0.015 \\
\hline
$\Gamma^{\rm tot}$& 0.019 & 0.017 \\
\hline
\end{tabular}
\label{tab:mesons4}
\end{table}

The construction of a meson exchange model was partly
motivated by the unability of the OPE mechanism to explain the
experimentally large neutron to proton ratio. A summary of
results for $\Gamma_n/\Gamma_p$ is shown in Table
\ref{tab:mesons5}. The spectacular increase in this ratio,
especially  for \hyp{12}{C}, obtained in \refer{holstein} when
all the mesons are included is not
confirmed by the results of \refer{assum97}.
It could be that the use of form factors, considered in
\refer{assum97}
and neglected in \refer{holstein}, diminish the effect of the
heavier mesons, due to the removal of strength at high momentum
values and thus at short distances, which would then produce a
ratio $\Gamma_n/\Gamma_p$ largely dominated by the OPE value.
In any case, the difference between the nuclear matter and
\hyp{12}{C} results in \refer{holstein} is somewhat surprising in
view of the fact that we are
dealing with a ratio of rates for which the short distances play
the most relevant role.
Details on the finite nucleus calculation of \refer{holstein} are
awaited in order to clarify the origin of such differences.

\begin{table}[hbt]
\centering
\caption{ Neutron- to proton-induced ratio $\Gamma_n/\Gamma_p$.
}
\begin{tabular}{|l c|c|c|c|}
\hline
 & & $\pi$ & $\pi,\rho$ & all mesons  \\
\hline
 & nuc. matter & 0.06 & 0.08 & 0.34 \\
\refer{holstein} & \hyp{5}{He} & 0.05 & & 0.48 \\
 & \hyp{12}{C} & 0.20 & & 0.83 \\
 \hline
 \refer{assum97} & \hyp{5}{He} & & & 0.07 \\
  & \hyp{12}{C} & 0.104 & 0.095 & 0.07 \\
\hline
\multicolumn{2}{|l|}{EXP:} & \multicolumn{3}{c|}{~} \\
 \cite{szymanski} & \hyp{5}{He} & \multicolumn{3}{c|}{$0.93 \pm
0.5$} \\
 & \hyp{12}{C} & \multicolumn{3}{c|}{$1.33^{+1.12}_{-0.81}$} \\
 \cite{noumi} & \hyp{12}{C} &
\multicolumn{3}{c|}{$1.87^{+0.91}_{-1.59}$} \\
 \cite{montwill} & \hyp{12}{C} & \multicolumn{3}{c|}{$0.70 \pm
0.30$} \\
 & & \multicolumn{3}{c|}{$0.52 \pm 0.16$} \\
\hline
\end{tabular}
\label{tab:mesons5}
\end{table}

The experimental values for $\Gamma_n/\Gamma_p$ have large
uncertainties due, mainly, to the difficulties in detecting
neutrons. Less uncertain are the results for the proton-induced
partial rate $\Gamma_p$ and the available data are compared in
Table \ref{tab:gammap} with the theoretical predictions of the
meson exchange model of \refer{assum97}.

\begin{table}[hbt]
\centering
\caption{ Proton-induced ratio $\Gamma_p$.
}
\begin{tabular}{|l|c|c|c|}
\hline
 & EXP & \multicolumn{2}{c|}{\refer{assum97}}  \\
 & & \multicolumn{1}{c}{$\pi$-only} & \multicolumn{1}{c|}{all
mesons} \\
\hline
\hyp{4}{He} & $0.16\pm 0.02$ \cite{zeps97} &  &\\
\hyp{5}{He} & $0.21 \pm 0.07$ \cite{szymanski} & 0.43 & 0.38 \\
\hyp{11}{B} & $0.30^{+0.15}_{-0.11}$ \cite{szymanski} & 0.64 &
0.56 \\
\hyp{12}{C} & $0.31^{+0.18}_{-0.11}$ \cite{szymanski} & 0.80 &
0.71 \\
 \hline
\end{tabular}
\label{tab:gammap}
\end{table}

Obviously, the fact that
the non-mesonic rate is well reproduced by this model while the
ratio $\Gamma_n/\Gamma_p$ turns out to be too small, leads
necessarily to an overestimation (of almost a factor of 2) for
$\Gamma_p$. It is somewhat surprising that, while both
neutron- and proton-induced partial rates appear to be in
disagreement with the data, their sum conspires to give a
non-mesonic rate which reproduces the measurements. Although a
more proper comparison should be made after the experimental cut
offs in the energy of the detected particles are implemented in
the theoretical models, it is important to stress that precise
measurements of $\Gamma_p$, more feasible experimentally than the
ratio $\Gamma_n/\Gamma_p$, would bring valuable information to
better understand the mechanism inducing the \lnnn transition.

\vskip 0.2cm

{\bf c) Two pion exchange}
\addcontentsline{toc}{subsubsection}{\protect\numberline{~~~{
c)}}{Two pion exchange}}

\vskip 0.2cm

\begin{figure}[htb]
       \setlength{\unitlength}{1mm}
       \begin{picture}(100,70)
      \put(25,-52){\epsfxsize=11cm \epsfbox{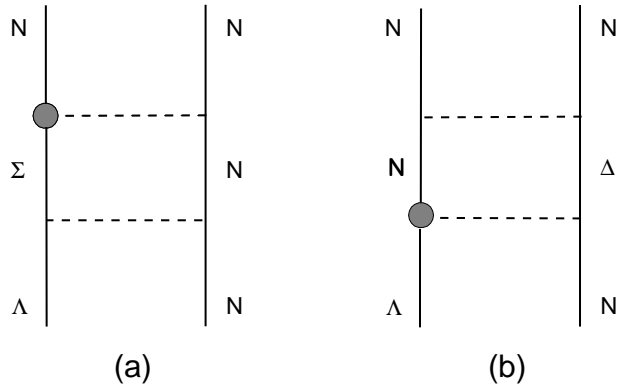}}
       \end{picture}
\caption{ Two pion exchange amplitudes to the \lnnn transition
through coupling to intermediate $\Sigma N$ (a) or $\Delta N$ (b)
states.
}
\label{fig:twopi}
\end{figure}

The two pion exchange mechanism of fig. \ref{fig:twopi}(a),
containing a
strong $\Lambda N \to \Sigma N$ transition plus a weak $\Sigma N
\to NN$ one, was studied by Band\={o} et al.  \cite{bando88}
together with the usual one pion mechanism for \lnnn. The
empirical $\Lambda N \pi$ and $\Sigma N\pi$ weak vertices were
used
and the $\Delta I=1/2$ rule enforced. A closure approximation was
employed to deal with the sum over intermediate states. The
potential for the $\Lambda
N \to \Sigma N \to N N$ transition was obtained for several
partial waves and compared with the OPE ones.
Important interference effecs for the final $I=0$ channels (only
accessible to $\Lambda p \to n p$), constructive for
$^3S_1\to^3S_1$  and $^3S_1 \to ^1P_1$ and destructive for
$^3S_1\to
^3D_1$, were found. As a result the calculated proton-induced
rate, $\Gamma_p$, for \hyp{5}{He} is larger than the OPE value in
about a factor 1.5. Moreover, the destructive interference in the
$I=1$ $^3S_1 \to ^3P_1$ channel produces an overall decrease in
the
neutron-induced rate, $\Gamma_n$. Therefore, the $\Lambda-
\Sigma$ coupling reduces the ratio
$\Gamma_n/\Gamma_p$ even more, from the OPE value 0.13 to 0.04.

In \refer{shmat94a}, this $\Lambda - \Sigma$ process was
considered together with the mechanism $\Lambda N \to N \Delta
\to NN$ of fig. \ref{fig:twopi}(b). The loop in figs.
\ref{fig:twopi} was computed explicitly and initial $\Lambda N$
and final $NN$ wavefunctions, correlated by the effect of the
strong interaction, were used. Note that a potential source of
doublecounting in the $\Lambda - \Sigma$ diagram may arise
because realistic $\Lambda N$ correlated wave functions have the
$\Lambda N \to \Sigma N$ transition built inside. The results for
the $\Lambda - \Sigma$ diagram are qualitatively different than
those of ref.  \cite{bando88} since a reduction of 30\% is found
in both the $^3S_1 \to ^3S_1$ and $^3S_1 \to ^3D_1$ amplitudes.
The $\Delta - N$ $^1 S_0 \to ^1S_0$ amplitude is about a factor
1.5 larger than the OPE one and would give rise to an increase on
$\Gamma_n$, although no rates, only amplitudes, were calculated
in \refer{shmat94a}.

\begin{figure}[htb]
       \setlength{\unitlength}{1mm}
       \begin{picture}(80,70)
      \put(22,-40){\epsfxsize=12cm \epsfbox{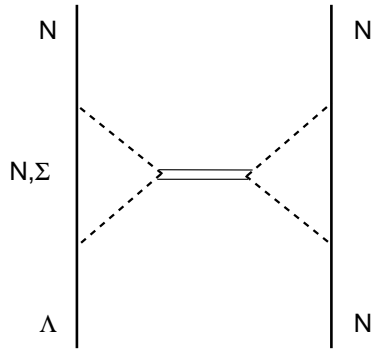}}
       \end{picture}
\caption{
Correlated two pion exchange \lnnn transition, where the pions
couple to $\sigma$ or $\rho$ mesons.
}
\label{fig:twopico}
\end{figure}

In analogy with recent developments in the $NN$ potential, in
which the $\rho$ and $\sigma$ mesons are considered in terms of
the exchange of two correlated pions, there have been recent
attemps to incorporate this picture in the weak \lnnn transition
potential \cite{shmat94b,itonaga}. The $\pi \pi$ correlation is
approximated, as displayed in fig. \ref{fig:twopico},
by a t-channel resonance which can be either the
scalar
$\sigma$ meson or the vector $\rho$ meson. The weak vertex is
obtained in both cases by means of a loop diagram with two pions
and an intermediate $N$ or $\Sigma$ baryon. The strong vertex is
generated in the same way in \refer{shmat94b}, while
phenomenological $\sigma NN$ and $\rho NN$ couplings from
ordinary one meson exchange potentials that fit the $NN$
scattering data are considered in \refer{itonaga}.
Note that the strength of the 2$\pi$ contribution obviously
depends on the cut off used to regularize the integrals in the
triangle loops.

In \refer{shmat94b} a substantial contribution of the 2$\pi$
diagram to the $^1 S_0 \to ^1S_0$ amplitude, about 4 times larger
than the OPE one, is found. For the other channels, the 2$\pi$
contribution represents at most 30\% of the $\pi$-only value.
This is in contrast to what is found in \refer{itonaga}, where
the 2$\pi/\rho$ potential in the $^ 3S_1 \to ^3 D_1$ channel
tends to cancel quite strongly the $\pi$-only contribution. The
$^1 S_0 \to ^1 S_0$ transition is also quite affected by the
2$\pi/\rho$ and 2$\pi/\sigma$ contributions, which have opposite
sign than that for the $\pi$-only one, but not at the level of
the factor 4 found in \refer{shmat94b}.

The model of \refer{itonaga} is applied to compute the partial
non-mesonic decay rates of \hyp{12}{C} \cite{iton97}. It is found
that the 2$\pi/\rho$ amplitudes decrease the non-mesonic rate
from
0.86 to 0.79 and including also the 2$\pi/\sigma$ ones gives a
final value of 0.92. The ratio $\Gamma_n/\Gamma_p$ goes from the
value 0.087 ($\pi$ only) to 0.10 ($\pi + 2\pi/\rho$) and to a
final 0.14 ($\pi + 2\pi/\rho + 2\pi/\sigma$), still far away from
the empirical results.

\subsubsection{Quark model based results}

In recent works \cite{oka,maltman} an old idea of treating the
\lnnn mechanism from the point of view of quark degrees of
freedom \cite{kisslinger} is reexamined. In the latter reference
a hybrid quark-hadron model was used in which the usual OPE
between
hadrons was adopted for the external region of relative $\Lambda
N$ distances ($r > r_0$ with $r_0 \simeq 0.8$ fm), while a
six-quark wave function and an effective quark Hamiltonian is
taken
for $r_0 < 0.8$ fm. The main uncertainty within that model is the
parametrization of the weak effective $\Delta S=1$ quark
Hamiltonian. Only when the coefficients of the QCD corrected
Gilman and Wise Hamiltonian \cite{gilman} are modified to
implement the $\Delta I= 1/2$ rule, did the non-mesonic decay of
\hyp{12}{C} fall within the uncertainties of the experimental
Brookhaven value of $1.14\pm 0.2$ \cite{szymanski}.

In refs.  \cite{oka,maltman} a four-quark $\Delta S=1$ effective
Hamiltonian, containing both $\Delta I=1/2$ and $3/2$ terms and
modified by QCD effects \cite{paschos} was used, along with the
quark cluster model for the quark component of the two-baryon
initial and final states. Since quark-antiquark pairs are not
allowed as intermediate states, the authors incorporate
explicitly meson exchange diagrams, as the $\pi$ meson \cite{oka}
and also the $K$ meson \cite{maltman}.

Both works find large $\Delta I=3/2$ contributions to the $^1S_0
\to ^1S_0$ and $^1S_0 \to ^3P_0$ amplitudes. However, it would
be desirable to clarify the origin of some sign and size
differences for certain amplitudes. For instance, the
neutron-induced contribution to the $^1 S_0 \to ^1 S_0$ amplitude
is $-0.07$ times the proton-induced one in \refer{oka}, while it
is a factor $-0.75$ the proton-induced one in \refer{maltman}.
The proton-induced $^1S_0$ $\to$ $^3P_0$ amplitude is much larger
in size in \refer{maltman} than the one obtained in \refer{oka}.
Therefore, although the results of both works are qualitatively
consistent, they still show important quantitative differences.


\begin{table}[hbt]
\centering
\caption{ Quark-model results of
ref. \protect\cite{oka2}
for the non-mesonic decay rate and
ratio $\Gamma_n/\Gamma_p$ of light hypernuclei. 
}
\begin{tabular}{|l|c|c|cc|}
\hline
 & OPE & OPE and DQ  &
 \multicolumn{2}{c|}{EXP} \\
\hline
$ \Gamma_{nm}$(\hyp{5}{He})& 0.216 & 0.627 & 
$0.41\pm 0.14$  \cite{szymanski} &  \\
$ \Gamma_{n}/\Gamma_p$(\hyp{5}{He})& 0.132 & 0.489 & 
$0.93\pm 0.55$  \cite{szymanski} &  \\
\hline
$ \Gamma_{nm}$(\hyp{4}{He})& 0.154 & 0.253 & 
$0.20\pm 0.05$  \cite{zeps97} &
$0.17\pm 0.05$  \cite{outa97}  \\
$ \Gamma_n/\Gamma_p$(\hyp{4}{He})& 0.061 & 0.178 & 
$0.25\pm 0.13$  \cite{zeps97} &
$0.06\pm 0.30$  \cite{outa97}  \\
\hline
\end{tabular}
\label{tab:quark2}
\end{table}

The amplitudes obtained in \refer{oka} were used to determine
the partial non-mesonic decay widths of s-shell nuclei but the
results depended on the relative sign between the quark
and the OPE contributions. 
Using the soft pion theorem,
the authors clarified recently \cite{oka2} the sign ambiguity.
This theorem allows one to relate the sign of the $\Lambda
\to n \pi^0$ coupling constant (relevant for the OPE mechanism)
to the PC baryon-baryon transition amplitude $\Lambda \to n$,
which was computed using the quark Hamiltonian and their quark
model wave functions. A more realistic $\Lambda$ wave function,
pushed to the surface by the effect of a repulsive
$\Lambda$-nucleus potential at short distances, was also used in
\refer{oka2}. The results, summarized in Table
\ref{tab:quark2}, improve substantially those obtained in their
previous calculations\cite{oka}. These results are promising but, before
calculations of the decay of heavier hypernuclei are attempted
within this model, it would be desirable to establish the
connection between the effective quark Hamiltonian and the
empirical $\Lambda \to N \pi$ vertex.

\subsubsection{$\Delta I=1/2$ violation?}

\vskip 0.2cm
{\bf a) Phenomenology}
\addcontentsline{toc}{subsubsection}{\protect\numberline{~~~{
a)}}{Phenomenology}}
\vskip 0.2cm

Non-mesonic weak decays of light hypernuclei can be used to test
the validity of the phenomenological $\Delta I=1/2$ rule. One
must first observe that, since the $\Lambda N$ pair is in a
$L=0$ state, the possible transitions are those listed in Table
\ref{tab:isorule1}.

\begin{table}[hbt]
\centering
\caption{ Possible  \lnnn transitions starting from a $L=0$
$\Lambda N$ pair.
}
\vskip 0.2cm

\begin{tabular}{cccc}
$^1S_0$ & $\to$ & $^1S_0$ & (I=1) \\
        & $\to$ & $^3P_0$ & (I=1) \\
$^3S_1$ & $\to$ & $^3S_1$ & (I=0) \\
        & $\to$ & $^3D_1$ & (I=0) \\
        & $\to$ & $^1P_1$ & (I=0) \\
        & $\to$ & $^3P_1$ & (I=1) \\
\end{tabular}
\label{tab:isorule1}
\end{table}

\vskip 0.5cm

The neutron-induced transitions can only lead
to $I=1$ final states and, if the $\Delta I=1/2$ rule applies,
then the ratio of neutron- to proton-stimulated transitions to
final $I=1$ states must be \cite{dalitz}
\begin{equation}
\frac{\Gamma_n^{(I=1)}}{\Gamma_p^{(I=1)}} = 2 \ .
\end{equation}

Therefore, by isolating the initial singlet spin state $^1S_0$,
which leads to only $I=1$ final states, both in neutron- and
proton-induced decays, the validity of the $\Delta I=1/2$ rule in
the
non-mesonic decay could be verified. This idea stimulated a large
amount of experimental
effort \cite{outa97,zeps97,schumacher,wein95}.
The connection
with experiments is done following the phenomenological model of
Block and Dalitz \cite{dalitz} and updated by Dover \cite{dover},
according to which the non-mesonic decay rates of light
hypernuclei
are expressed in terms of elementary interaction strengths
between
singlet or triplet $\Lambda N$ pairs:
\begin{eqnarray}
\Gamma_{nm}({}^4_\Lambda{\rm H}) & = & \frac{1}{6} \rho_3 ( 3
R_{n1} +
\phantom{2}R_{n0} \phantom{ + 3 R_{p1} \;} + 2 R_{p0})
\label{eq:h4}\\
\Gamma_{nm}({}^4_\Lambda{\rm He}) & = & \frac{1}{6} \rho_3 (
\phantom{ 3
R_{n1} + \;} 2 R_{n0} + 3 R_{p1} + \phantom{2} R_{p0})
\label{eq:he4} \\
\Gamma_{nm}({}^5_\Lambda{\rm He}) & = & \frac{1}{8} \rho_4 ( 3
R_{n1} +
\phantom{2}R_{n0} + 3 R_{p1} + \phantom{2} R_{p0}) \ .
\label{eq:he5}
\end{eqnarray}
The quantities $R_{NS}$ stand for the non-mesonic decay of a
$\Lambda N$ pair with spin $S$, for unit density of nucleon $N$
at the $\Lambda$ position, and $\rho_A$ denotes the mean nucleon
density at the $\Lambda$ position.
An average over spin and charge is implied in
eqs. (\ref{eq:h4})$-$(\ref{eq:he5}). These expressions also assume
that the final state interactions, which were shown to be
important in the non-mesonic decay of the hypertriton \cite{golak97},
are contained in the elementary interaction strengths and thus 
are the same in all these light hypernuclei. 
The densities cancel in forming the ratios
\begin{eqnarray}
\gamma^4 & = & \frac{\Gamma_n({}^4_\Lambda{\rm He})}
{\Gamma_p({}^4_\Lambda{\rm He})} = \frac{2 R_{n0}}{3
R_{p1}+R_{p0} }
\\
\gamma^5 & = & \frac{\Gamma_n({}^5_\Lambda{\rm He})}
{\Gamma_p({}^5_\Lambda{\rm He})} = \frac{3 R_{n1} + R_{n0}}{3
R_{p1}+
R_{p0} } \\
\gamma_{nm} & = & \frac{\Gamma_{nm}({}^4_\Lambda{\rm He})}
{\Gamma_{nm}({}^4_\Lambda{\rm H})} = \frac{2 R_{n0} + 3 R_{p1} +
R_{p0}}
{3 R_{n1}+ R_{n0} + 2 R_{p0} } \ ,
\end{eqnarray}
all of which involve experimentally known decay rates, as seen in
Table \ref{tab:isorule2}, and therefore allow one to extract the
ratio
\begin{equation}
\kappa = \frac{R_{n0}}{R_{p0}} = \frac{\gamma_{nm} \gamma^4}{1 +
\gamma^4 - \gamma_{nm} \gamma^5} \ .
\end{equation}

\begin{table}[hbt]
\centering
\caption{ Experimental weak decay observables of several
s-shell hypernuclei.
}
\begin{tabular}{|l|c|cc|c|}
\hline
 & \hyp{5}{He} & \multicolumn{2}{c|}{\hyp{4}{He}} & \hyp{4}{H} \\
 &  \cite{szymanski} &  \cite{zeps97} &  \cite{outa97} &
 \cite{outa97} \\
\hline
$ \Gamma_{n}$ & $0.20 \pm 0.11$  & $0.04 \pm 0.02$ & $0.01 \pm
0.05$ &  \\
$ \Gamma_{p}$ & $0.21 \pm 0.07$  & $0.16 \pm 0.02$ & $0.16 \pm
0.02$ &  \\
$ \Gamma_{n}/\Gamma_p$ & $0.93 \pm 0.55$ & $0.25 \pm 0.13$ &
$0.06 \pm 0.30$ &  \\
$ \Gamma_{nm}$ & $0.41 \pm 0.14$ &
$0.20\pm 0.05$ &  $0.17\pm 0.05$ &
$0.17\pm 0.11$   \\
\hline
\end{tabular}
\label{tab:isorule2}
\end{table}

Taking the central values of the KEK \cite{outa97} or
BNL \cite{zeps97}
experiments, the value for $\kappa$ is 0.4 or 0.8 but the error
bars are so large that a value of 2, compatible with the $\Delta
I=1/2$ rule, is not ruled out \cite{wein95}. A much more direct
determination of $\kappa$ would be obtained from
\begin{equation}
\kappa=\frac{\Gamma_n({}^4_\Lambda{\rm
He})}{\Gamma_p({}^4_\Lambda{\rm H}
)} =
\frac{R_{n0}}{R_{p0}} \ ,
\end{equation}
which requires the measurement of
the partial rates for \hyp{4}{H}, an attempt that will be made in
experiment E907 at BNL \cite{zeps97}.
Since the partial neutron rate for \hyp{4}{He} has been
measured to be small, even if the error bar is large, the $\Delta
I=1/2$ rule would pose severe constraints on the value of
$\Gamma_p$(\hyp{4}{H}), which could not exceed a few percent of
the free $\Lambda$ decay rate.

\vskip 0.2cm
{\bf b) Models}
\addcontentsline{toc}{subsubsection}{\protect\numberline{~~~{
b)}}{Models}}
\vskip 0.2cm

 The quark model of refs.  \cite{oka,maltman} is based on the
effective 4 quark Hamiltonian with QCD corrections \cite{gilman}
extended down to an energy scale $\mu^2\simeq\mu_0^2$
 \cite{paschos} at which $\alpha(\mu_0^2)=1$. This quark
Hamiltonian contains some enhancement of the $\Delta I=1/2$
transitions although it is still insufficient to explain the
ratio of $\Delta
I=1/2$ to $\Delta I=3/2$ amplitudes for the non-leptonic weak
decay of $K$, $\Lambda$ and other strange hadrons. As mentioned
in the previous section, two body amplitudes were evaluated by
combining the quark Hamiltonian with a constituent quark model
for the baryons. In spite of the numerical differences, both
models predict substantial $\Delta I =3/2$ contributions to the
spin singlet amplitudes $^1 S_0 \to ^1S_0$ and $^1S_0 \to ^3P_0$.
Therefore, these are the most important ones to measure in order
to determine violations of the $\Delta I=1/2$ rule.

In \refer{assum32} violation of the $\Delta I=1/2$ rule in the
\lnnn mechanism was
investigated in the framework of a meson exchange model. The work
is based on the observation that, while the $\Delta I=3/2$
contributions to the weak coupling constants of baryons to the
pseudoscalar mesons are empirically known to be small, this is
not necessarily the case for the vector mesons. Indeed, based on
the weak QCD-corrected four quark effective Hamiltonian at a
scale
$\simeq$ 1 GeV \cite{donoghue2}, the factorization contributions
to the weak $\Lambda N \rho$ and $\Sigma N \rho$ couplings were
found to contain $\Delta I=3/2$ terms comparable in magnitude to
the $\Delta I=1/2$ ones \cite{maltman95a}. The
factorization
approximation in the evaluation of the $\Delta I=3/2$ amplitudes
for the known hyperon decays yields \cite{maltman95b}: (1) good
fits for the s-wave and p-wave $\Xi$ amplitudes and the s-wave
triangle discrepancy, (2) an underestimate of the p-wave $\Sigma$
triangle discrepancy by a factor 3--4, and (3) an overestimate of
the s-wave and p-wave $\Lambda$ amplitudes by a factor 3--4. It
is important to notice, however, that the experimental errors on
the $\Delta I=3/2$ amplitudes are large and the above mentioned
discrepancies are within the 2$\sigma$ level. The work of
\refer{assum32} takes the predicted $\Delta I=3/2$ contributions
to the coupling constants of the vector mesons and rescales them
by
factors ranging in between $(-3,3)$ to account for the
limitations of the factorization model. Their $\Delta I=1/2$
potential was extended to incorporate the $\Delta I=3/2$
transitions of the $\rho$ and $K^*$ mesons by assuming the
$\Lambda$ to behave as a $|3/2\ -1/2 >$ isospurion and
replacing the isospin $\vec{\tau} \vec{\tau}$ operator in the
potential by $\vec{\tau}_{3/2} \vec{\tau}$, where
$\vec{\tau}_{3/2}$ is the usual isospin $3/2 \to 1/2$ transition
operator.
Even if the $\Delta I=3/2$ couplings of the vector mesons were
comparable in magnitude to the corresponding $\Delta
I=1/2$ ones, it was found that, when all the other mesons were
incorporated, the decay rate of \hyp{12}{C} changed at most by
6\%, the proton-induced rate was affected at most by 10\%, while
the neutron-induced rate could change by up to a factor 2. In the
most favorable scaling situation (a scale factor 3 for both
$\rho$
and $K^*$) the ratio $\Gamma_n/\Gamma_p$ changed from the $\Delta
I=1/2$ value of 0.07 to 0.14, which is still away from the
experimental values.

These findings are substantially different than those found in
the
quark model works \cite{oka,maltman}, were the $\Delta I=3/2$
amplitudes
seem to play a much more relevant role. Therefore a direct
determination of the spin singlet ratio $R_{n0}/R_{p0}$, through
the measurement of nucleon induced decays of light hypernuclei,
is
very much desired in order to clarify the issue of $\Delta I=1/2$
violation in the non-mesonic weak decay.

\subsection{Asymmetry}

The possibility of measuring the asymmetry in the angular
distribution of protons from the weak decay of polarized
hypernuclei offers a unique opportunity to learn about the weak
decay mechanism. The asymmetry is due to an interference between
PC and PV amplitudes and, hence, can give complementary
information to that obtained from total and partial rates which
are dominated by the PC piece of the \lnnn interaction.

The first observation of asymmetries on the decay products from
polarized hypernuclei was done at KEK \cite{ajim}.
There, polarized hypernuclei (${}^{12}_\Lambda\vec{\rm C}$,
${}^{11}_\Lambda\vec{\rm B}$) were created through the
$(\pi^+,K^+)$
reaction, which has been shown to produce nonnegligible
polarization perpendicular to the reaction plane for a momentum
of the pion $p_\pi \sim 1.04$ GeV/c at small $K^+$ scattering
angles ($10-15^\circ$) where the cross section is still
appreciable \cite{bando89,iton94}.

The intensity of protons emitted at an angle $\chi$ with respect
to the polarization axis in the decay of polarized hypernuclei is
given by
\begin{equation}
I(\chi) = {\rm Tr} ({\cal M} \rho {\cal M}^\dagger ) \ ,
\label{eq:asym1}
\end{equation}
where ${\cal M}$ is the hypernuclear transition operator and
\begin{equation}
\rho(J) = \frac{1}{2J + 1} \left( 1 + \frac{3}{J+1} P_y S_y
\right)
\label{eq:dens}
\end{equation}
is the density matrix describing the hypernucleus of spin
$J$ polarized along the $y$ axis, which is perpendicular to the
reaction plane. Eq. (\ref{eq:asym1}) can be written as
\begin{equation}
I(\chi)=I_0 \left(1+ P_y A_y(\chi) \right) \ ,
\end{equation}
where $I_0$ is the isotropic intensity of protons from the decay
of unpolarized hypernuclei and $A_y(\chi)$ the asymmetry
parameter which reads
\begin{equation}
A_y(\chi) = \frac{3}{J+1} \frac{ {\rm Tr} ({\cal M} S_y
{\cal M}^\dagger  ) }{ {\rm Tr} ({\cal M} {\cal M}^\dagger ) } =
\frac{3}{J+1} \frac{ \displaystyle\sum_{M_i} \sigma(M_i)
M_i}{\displaystyle\sum_{M_i}
\sigma(M_i) } \cos{\chi} = A_y \cos{\chi}
\label{eq:asym2}
\end{equation}
The last equation shows that the asymmetry exhibits a simple
$\cos{\chi}$ dependence and is given in terms of $\sigma(M_i)$,
the weak decay probability for a hypernucleus with spin
projection $M_i$ ejecting protons along the quantization
axis. The experiment compares the total number of protons
emerging parallel or antiparallel to the $y$-axis and ,
therefore, the product $P_y A_y$ is determined. Using standard
angular momentum algebra and assuming the $\Lambda$ to couple
only to the nuclear core ground state of spin $J_c$, one can
obtain the $\Lambda$ polarization as
\begin{equation}
p_\Lambda = \left\{ \begin{array}{cl}
-\displaystyle\frac{J}{J+1} P_y & ~~~{\rm if}\ \ J=J_c-1/2 \\
P_y & ~~~{\rm if}\ \ J=J_c+1/2
\end{array} \right.
\end{equation}
and define the intrinsic $\Lambda$ asymmetry parameter
\begin{equation}
a_\Lambda = \left\{ \begin{array}{cl}
-\displaystyle\frac{J+1}{J} A_y & ~~~{\rm if}\ \ J=J_c-1/2 \\
A_y & ~~~{\rm if}\ \  J=J_c+1/2
\end{array} \right.
\end{equation}
such that the asymmetry
\begin{equation}
{\cal A} = P_y A_y \cos{\chi} = p_\Lambda a_\Lambda \cos{\chi}
\label{eq:asym3}
\end{equation}
is expressed in terms of $\Lambda$ properties, $p_\Lambda$ and
$a_\Lambda$, which are then
characteristic of the elementary $\vec{\Lambda} N \to NN$
reaction.

In order to compare with experiment, the polarization
of the hypernucleus must be known. Until now
this information is not available and it is necessary to resort
to theoretical models, which take into account the conditions of
the production reaction. Given the energy resolution of the KEK
experiment \cite{ajim} ($5-7$ MeV), several hypernuclear
states can be excited, which then decay electromagnetically
and/or
by particle emission to a hypernuclear ground state prior
to the weak decay. To determine the polarization at
this stage one requires: i) the amount of polarization of the
states excited by the production mechanism, along with their
corresponding formation cross sections, and ii) an attenuation
coefficient to account for the loss of polarization of each level
in the deexcitation process \cite{ejiri}.

In \refer{bando89} one-particle one-hole ($j_n^{-1} j_\Lambda$)
hypernuclear wave
functions were employed to obtain the polarizations of typical
states within the p-shell region. The model was
refined in \cite{iton94} by incorporating configuration-mixed
shell-model wave functions.
Using their predictions for the three $1^-$ states in \hyp{12}{C}
and the depolarization formalism of \refer{ejiri}, one obtains
$P_y=-0.19$ ($p_\Lambda = 0.095$).
On the other hand, the hypernucleus $^{11}_\Lambda$B is created
by particle emission
from
excited states of $^{12}_\Lambda$C. The
window of excitation energy that spans 1.55 MeV between the
(p $ + ^{11}_\Lambda$B) and the ($\Lambda + ^{11}$C) particle
decay
threshold contains three positive-parity states: two 2$^+$ states
separated by $\sim$ 800 KeV and a narrow 0$^+$ state just below
the
($\Lambda + ^{11}$C) threshold.
Using the model of ref. \cite{iton94},
which predicts equal formation cross sections for
the $2^+_1$ and $2^+_2$ states, and neglecting the 0$^+$ state
because
of its relatively small cross section,
a polarization of
$P_y=-0.29$ ($p_\Lambda=0.21$) is obtained. However, hypernuclear
structure calculations by
Auerbach et al.  \cite{auerbach} predicted strong
configuration
mixing which
reduced the cross section of the lower $2^+$ state by a factor of
three relative to the higher one. This prediction was verified by
a
reanalysis of older emulsion data \cite{dalitz2}. Taking these
relative
weights into account, one obtains the more realistic value
$P_y=-0.43$ ($p_\Lambda=0.31$).

In Table \ref{tab:asym} a comparison of theoretical results to
the presently available data is made. The two meson exchange
models predict a quite similar asymmetry parameter $a_\Lambda$
which ranges between $-0.3$ and $-0.4$ for the different
hypernuclei.
The comparison with experiment must be made at the level of the
asymmetry, for which it is necessary to multiply $a_\Lambda$ by
the $\Lambda$ polarization discussed above. Good agreement with
the data is obtained. However, the error bars are still so large
that not much can be learned about the mechanism governing the
decay.

\begin{table}[hbt]
\centering
\caption{ Proton asymmetry from the weak decay of hypernuclei.
}
\begin{tabular}{|l|ccc|c|}
\hline
 & \multicolumn{3}{c|}{Parre\~no et al.  \cite{assum97}} & Dubach
et al.  \cite{holstein} \\
 & \hyp{5}{He} & \hyp{11}{B} & \hyp{12}{C} & nuc. matter \\
\hline
$ a_{\Lambda}$ ($\pi$ corr.) & & & $-0.238$ & $-0.192$   \\
$ a_{\Lambda}$ (all mesons) & $-0.273$ & $-0.391$ & $-0.316$ & $-
0.443$   \\
\hline
$ p_\Lambda$ & & 0.31 & 0.095 & \\
\hline
${\cal A}=p_\Lambda a_\Lambda$ & & $-0.12$ & $-0.13$ & \\
\hline
${\cal A}$ (EXP) \cite{ajim} & & $-0.20 \pm 0.10$ & $-0.01 \pm
0.10$ & \\
\hline
\end{tabular}
\label{tab:asym}
\end{table}

In order to avoid the need for theoretical input and access $A_y$
directly, a new experiment at KEK \cite{kishi} was devised to
measure
the decay of polarized $^{5}_\Lambda$He, extracting both
the pion asymmetry from the mesonic
channel, ${\cal A}_{\pi^-}$, and the proton asymmetry from the
non-mesonic decay, ${\cal A}$.
The asymmetry parameter $a_{\pi^-}$ of the pionic channel has
been estimated to be very similar to that of the free $\Lambda$
decay \cite{motoba}, and, therefore, the hypernuclear
polarization
can be
obtained from the relation $P_y={\cal A}_{\pi^-}/a_{\pi^-}$.
Combining this value with the measured proton asymmetry
${\cal A}$, permits to determine the
asymmetry parameter for the non-mesonic decay from the equality
$A_y={\cal A}/P_y$.

A preliminar analysis of this experiment has recently become
available \cite{ajim97} and gives:
$p_\Lambda = 0.217 \pm 0.087 \pm 0.021$ for an emerging $K^+$
angle $2 < \mid \theta_K \mid < 7^\circ $ and
$p_\Lambda = 0.382 \pm 0.114 \pm 0.013$ for
$7 < \mid \theta_K \mid < 15^\circ$. For the proton asymmetry the
measured values are:
${\cal A} = 0.102 \pm 0.050 \pm 0.005$ for
$2 < \mid \theta_K \mid < 7^\circ$ and
${\cal A} = 0.045 \pm 0.099 \pm 0.021$ for
$7 < \mid \theta_K \mid < 15^\circ$.
The combination of these preliminary results, which show large
uncertainties, would give rise to a positive asymmetry parameter
$a_\Lambda$, in total contradiction with their previous
results \cite{ajim} and the theoretical
predictions shown in Table \ref{tab:asym}. One would also expect
a
straight line, passing through the origin, when
representing the asymmetry ${\cal A}=p_\Lambda a_\Lambda$ as a
function of $p_\Lambda$. The two new experimental points do not
even show this behavior. However, one should keep in mind that
corrections due to the leaking of pions into the proton
identification area still need to be made \cite{ajim97} and,
therefore, the analysis is not yet complete.

\subsection{Two-nucleon induced decay}

\begin{figure}[htb]
       \setlength{\unitlength}{1mm}
       \begin{picture}(100,70)
      \put(30,-45){\epsfxsize=11cm \epsfbox{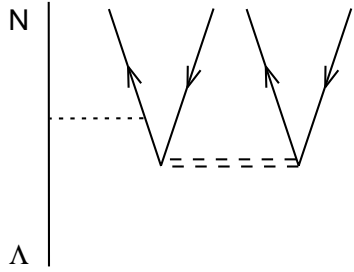}}
       \end{picture}
\caption{
Schematic representation of the $\Lambda$ decay coupling to
2p2h
components through virtual (close to real) pion absorption.
}
\label{fig:twon}
\end{figure}

The relevance of the two-nucleon induced channel, $\Lambda N N
\to N N N$, in the decay of hypernuclei was first pointed out by
Alberico et al. \cite{alberico}. This transition can be viewed as
coming from the absorption of a virtual pion emitted in the weak
vertex by two strongly correlated nucleons, as depicted
schematically in fig. \ref{fig:twon}. An interesting point raised
in \refer{alberico} was that this channel affects the emission of
neutrons and could have some relevance in the experimental
determination of the ratio $\Gamma_n/\Gamma_p$. Since pions are
absorbed mainly by neutron-proton pairs, the $2N$-induced decay
proceeds basically through the mode $\Lambda np \to nnp$,
emitting twice as many neutrons than protons. Hence, an
experimental observation of a large number of neutrons could
still be compatible with a smaller value for $\Gamma_n/\Gamma_p$
if the $2N$ mechanism was appreciably large.

In \refer{alberico} the decay width of a $\Lambda$ in nuclear
matter was calculated within the propagator method formalism
discussed in Sect. 2.1.1, which allows treating all the decay
channels in an unified way. Eq. (10) needs to
be modified to include the effect of form factors and short-range
correlations, which have been shown to be important for the
$1N$-induced mechanism \lnnn.
Thus, a monopole form factor at both the strong and weak vertices
\begin{equation}
F(q)=\frac{\Lambda^2-\mu^2}{\Lambda^2-q^2} \ ,
\end{equation}
with $\Lambda \sim 1.2$ GeV is included. Moreover, the pion
interaction between two nucleons is replaced by the
particle-hole interaction of refs.  \cite{oset79,oset82}
\begin{equation}
V_{ph}=\left\{ V_L(q) \hat{q}_i \hat{q}_j + V_T(q) (\delta_{ij} -
\hat{q}_i \hat{q}_j ) \right\} \sigma_i \sigma_j \vec{\tau}
\vec{\tau} \ ,
\end{equation}
which includes $\pi$ and $\rho$ exchange modulated by the
effect of short range correlations.
Detailed expressions of $V_L(q)$ and $V_T(q)$ can be found in
\refer{oset}. It is convenient to split $V_L(q)$ into the free
pion propagator plus the Landau-Migdal parameter,
$g^\prime$, which here is slightly momentum dependent
\begin{equation}
V_L(q)=\left( \frac{f}{\mu} \right)^2 \left[ \vec{q}\,^2 F^2(q)
D_0(q) + g^\prime(q) \right] \ .
\end{equation}
The $\Lambda N$ correlations are treated in a similar
way as in \cite{oset} by modifying the pion exchanged between the
weak
and strong vertices. The decay width can be finally written as
\begin{eqnarray}
\Gamma(k) & = & -6 ( G_F\mu^2)^2 \int \frac{d^3q}{(2\pi)^3} [1 -
n(\vec{k}-\vec{q}\,) ] \theta(k^0-E(\vec{k}-\vec{q}\,) - V_N)
\nonumber \\
           & \times & \left. {\rm Im}\, \alpha(q)
\right|_{q^0=k^0-
E(\vec{k}-\vec{q}\,) - V_N} \ ,
\label{eq:gamma2}
\end{eqnarray}
with
\begin{eqnarray}
\alpha(q) & = & \left( S^2 + \left(\frac{P}{\mu} \right)^2
\vec{q}\,^2 \right) F^2(q) D_0(q)  \nonumber \\
          & + & \frac{\tilde{S}^2(q) U(q)}{1-V_L(q) U(q)}
           +  \frac{\tilde{P}_L^2(q) U(q)}{1-V_L(q) U(q)}
\nonumber \\
          & + & 2 \frac{\tilde{P}_T^2(q) U(q)}{1-V_T(q) U(q)} \ ,
\label{eq:alpha}
\end{eqnarray}
where the functions $\tilde{S}(q)$, $\tilde{P}_L(q)$ and
$\tilde{P}_T(q)$, defined explicitly in \refer{oset}, reduce to
\begin{eqnarray}
\tilde{S}(q)& \to & \left(\frac{f}{\mu} \right) S D_0(q) F^2(q)\\
\tilde{P}_L(q) & \to & \left(\frac{f}{\mu} \right) \frac{P}{\mu}
\vec{q}\,^2 D_0(q) F^2(q)\\
\tilde{P}_T(q) & \to & 0
\end{eqnarray}
in the absence of $\Lambda N$ short range correlations.

The $2N$-induced $\Lambda$ decay mode appears when one considers
the absorption of the virtual pion by 2 particle-2 hole ($2p2h$)
states. Therefore, the function $U(q)$ in \eqn{eq:alpha} must
include a term, $U_{2p2h}(q)$, that accounts for this coupling
and becomes
\begin{equation}
U(q)=U_N(q)+U_\Delta(q) + U_{2p2h}(q) \ .
\end{equation}

Before going into the details on how this function is determined,
let us illustrate the physical mechanism that leads to the new
decay channel. The strength of a free pion, which is accumulated
in a delta function around a momentum value of 100 MeV/c,
spreads in the medium over a wide range of energies due to the
coupling of the pion to
$ph$, $\Delta h$ and $2p2h$ excitations. This distribution
presents a Breit-Wigner type peak, the position of which
changes slightly from the original pion pole.
The energy and momentum of the pion at the peak is such that, at
normal nuclear matter density, still enforces a nucleon momentum
which lies below the Fermi momentum.
However, the width of the
distribution is such that part of its tail corresponds to a Pauli
unblocked situation and, since at low pion energies this width is
mostly due to pion absorption through $2p2h$ states, as shown in
fig. \ref{fig:twon}, the new mode would be observed as three
particle emission from $\Lambda NN \to NNN$.

In \refer{alberico} the following parametrization was used
\begin{equation}
U_{2p2h} = -4\pi \vec{q}\,^2 \rho^2 {\cal C}_0 \ ,
\end{equation}
with a value ${\rm Im}\, {\cal C}_0= 0.18 \mu^{-6}$ taken from
theoretical studies performed in connection with $(e,e^\prime)$
reactions \cite{albe84} and compatible with data from pionic
atoms,
provided the real part of ${\cal C}_0$ was ${\rm Re}\, {\cal
C}_0=
0.40 \mu^{-6}$. The value of $U_{2p2h}$ was taken constant, even
if the phase space in the $\Lambda$ decay forces one to move away
from the situation in pionic atoms to values of $(q^0,\vec{q}\,)$
different from $(\mu,\vec{0})$. Moreover, ${\rm Im}\, U_{2p2h}$
was
set to zero in the region where ${\rm Im}\, U_N \neq 0$.
Their results for the $2N$-induced decay rate of a $\Lambda$ in
nuclear matter at an average nuclear density $\rho=0.69 \rho_0$,
where $\rho_0$ is the normal density, are shown in Table
\ref{tab:twon}, splitted into the s-wave, p-wave longitudinal and
p-wave transverse contributions. The final result of 0.52
represents a fraction of about 40\% of the $1N$-induced decay
rate.

\begin{table}[hbt]
\centering
\caption{ $2N$-induced contribution to the $\Lambda$ decay
rate in nuclear mattter.
}
\begin{tabular}{|l|c|c|c|}
\hline
 &  \cite{alberico} & ${\cal C}_0^*$ & Ph. space \\
\hline
$S$ & 0.40 & 0.119 & 0.148 \\
$P_L$ & 0.079 & 0.023 & 0.026 \\
$P_T$ & 0.040 & 0.023 & 0.065 \\
\hline
$\Gamma_{2p2h}$ & 0.52 & 0.165 & 0.238 \\
\hline
\end{tabular}
\label{tab:twon}
\end{table}

The relative large strength of this channel and the repercussion
it has in the proper interpretation of the ratio
$\Gamma_n/\Gamma_p$ motivated a more detailed
analysis \cite{ramos94}, in which a more realistic pion
self-energy
was used and the results were extended to finite nuclei via the
LDA. The determination of $U_{2p2h}$ was
based on new developments in the extraction of the optical
potential from pionic atoms \cite{meirav89}, reanalyzed in
\refer{garcia92} to account for different neutron and proton
radii, which yielded
\begin{equation}
\Pi_{2p2h}(q^0=\mu,\vec{q}\,\sim 0,\rho) = - 4\pi \vec{q}\,^2
\rho^2
\left(1 + \frac{1}{2} \varepsilon \right)^{-1} {\cal C}_0
\label{eq:proper}
\end{equation}
with
\begin{eqnarray*}
\varepsilon & = & \frac{\mu}{M} \\
{\cal C}_0 & = & (0.073 + i 0.068 ) \mu^{-6}
\end{eqnarray*}
However, since the propagator method model described above
generates the Lorentz-Lorenz correction automatically, one needs
to rewrite the information of eq. (\ref{eq:proper}) accordingly.
Thus, for the average density of pionic atoms in p-wave
($0.75\rho_0$ according to \cite{seki83}), the results of our
formulation should coincide with the experimental one given in
eq. (\ref{eq:proper}). This is readily obtained with the help of
eq. (8) and
the result for $U_{2p2h}$ is \cite{ramos94}
\begin{equation}
\left(\frac{f}{\mu}\right)^2 \vec{q}\,^2
U_{2p2h}(q^0=\mu,\vec{q}\,\sim
0,
\rho) = - 4\pi \vec{q}\,^2 \rho^2 {\cal C}_0^* \ ,
\end{equation}
with
\begin{equation}
{\cal C}_0^* = (0.105 + i 0.096) \mu^{-6} \ ,
\end{equation}
which yields a value for ${\rm Im}\, {\cal C}_0^*$ half of that
of \refer{alberico} and a value for ${\rm Re}\, {\cal C}_0^*$
four times
smaller. The extension of the self-energy to new kinematical
regions away from pionic atoms was also performed in
\refer{ramos94}.
The importance of all these modifications is displayed in Table
\ref{tab:twon}, where the third column corresponds to the same
calculation of \refer{alberico} but replacing their ${\cal C}_0$
value by the new ${\cal C}_0^*$ one. The results are reduced by a
factor 3 due to a smaller absorptive ${\rm Im}\, {\cal C}_0^*$
and
a less attractive ${\rm Re}\, \Pi_{2p2h}$ which moves the pion
pole
to a smaller momentum. Not forcing ${\rm Im}\, U_{2p2h} = 0$ in
the
$ph$ excitation region (where ${\rm Im}\, U_N \neq 0$) and taking
the
phase-space modifications into account increase the results
further, as seen in the last column of Table \ref{tab:twon}, but
the final value is less than half that obtained
previously \cite{alberico}.

Even if this decay channel turns out to be considerably smaller
than
claimed before, it will be shown in the next section that its
consideration has repercussions in the experimental determination
of the $\Gamma_n/\Gamma_p$ ratio. This investigation requires to
have predictions for the $2N$-induced decay rate of finite
nuclei, which
were also obtained in \refer{ramos94} on the basis of the
LDA shown in eq. (13). The results
are displayed in
Table \ref{tab:rates} for different hypernuclei, where, for
completeness, the decay rates from the mesonic and $1N$-induced
channels are also shown.

\begin{table}[htb]
\centering
\caption{ Mesonic, $1N$-induced and $2N$-induced decay rates
for several hypernuclei in the LDA \cite{ramos94}.
}
\begin{tabular}{|l|c|c|c|}
\hline
 & $\Gamma_{m}$ & $\Gamma_{nm}$ & $\Gamma_{2p2h}$ \\
\hline
\hyp{12}{C} & 0.31 & 1.45 & 0.27 \\
\hyp{16}{O} & 0.24 & 1.54 & 0.29 \\
\hyp{20}{Ne} & 0.14 & 1.60 & 0.32 \\
\hyp{40}{Ca} & 0.03 & 1.76 & 0.32 \\
\hyp{56}{Fe} & 0.01 & 1.82 & 0.32 \\
\hyp{89}{Y}  & ---  & 1.88 & 0.31 \\
\hyp{208}{Pb} & --- & 1.93 & 0.30 \\
\hline
\end{tabular}
\label{tab:rates}
\end{table}

The total lifetimes ---inverse of the total decay widths--- are
somewhat small compared with the recent lifetime measurements at
KEK \cite{bhang97}, which in the case of \hyp{56}{Fe} is $215 \pm
14$ ps, a factor 1.8 larger than the result obtained from Table
\ref{tab:rates}. However, the lifetime of \hyp{208}{Pb} derived
from the table is only 20\% smaller than the central value
$145\pm 7 \pm 23$ ps
obtained for \hyp{209}{Bi} in J\"ulich \cite{schult97} and
compatible within errors. The latter measurements are done by
bombarding heavy nuclei with protons and looking at the delayed
fission fragments. The method was proved to work in \refer{ohm97}
and used in the experiment of \refer{schult97} in order to obtain
the hypernuclear widths with more precission. These latter
experimental results are also compatible with those of
\refer{armstrong93} obtained from $\overline{p}$ interactions
with nuclei which, however, have large uncertainties.
One must note that the theoretical results are somewhat
dependent on the $\Lambda N$ correlations, controlled by the
$g^\prime_\Lambda$ parameter, and on the $\Lambda$ wave function
used in the LDA. The $1N$-induced
rate for \hyp{12}{C} would be reduced from 1.45 to 1.26 if a more
extended $\Lambda$ wave function with an oscillator parameter
$b_\Lambda = 1.87$ fm and a value $g^\prime_\Lambda=0.2$ were
used. This value for $g^\prime_\Lambda$ is what corresponds to a
Bessel
type correlation
function $f(r)=1-j_0(q_c r)$ with $q_c=3.93$ fm$^{-1}$, very
similar to the correlation function used in \refer{assum97} based
on realistic $\Lambda N$ wavefunctions obtained from the Nijmegen
interactions.
Therefore, the results in Table \ref{tab:rates} could easily be
10-15\%
smaller and, correspondingly, the lifetimes 10-15\% larger,
which would then agree better with the
very heavy hypernuclei results \cite{schult97,armstrong93} but
would still be in disagreement
with the KEK ones.

\subsection{The $\Gamma_n/\Gamma_p$ puzzle}

As mentioned in the previous section, one of the interesting
consequences of the $2N$-induced mechanism is that, since pions
are absorbed predominantly by $np$ pairs, this decay mechanism
would emit two neutrons and one proton. Hence, part of the large
number of observed neutrons would then come from the $2N$-induced
channel and the ratio $\Gamma_n/\Gamma_p$ would not need to be so
large as what is found in the present analyses, which associate
all
non-mesonic decay to the $1N$-induced mechanisms. Actually, as
shown in \refer{ramos94} and summarized below, the situation is
not so clear.

The experimental analysis of \refer{szymanski} takes only the
$1N$-mechanism into account and the assignment of the number of
neutrons and protons goes as
\begin{eqnarray}
N_n & \sim & 2 \Gamma_n^{\rm exp} + \Gamma_p^{\rm exp} \nonumber
\\
N_p & \sim & \Gamma_p^{\rm exp} \label{eq:old}
\end{eqnarray}
with  $\Gamma_{nm}^{\rm exp} = \Gamma_n^{\rm exp} +
\Gamma_p^{\rm exp}$.
However, considering the $2N$-induced mechanism and assuming that
all emitted nucleons are detected, the appropriate analysis reads
\begin{eqnarray}
N_n & \sim & 2 \Gamma_n + \Gamma_p + 2\Gamma_{2p2h} \nonumber \\
N_p & \sim & \Gamma_p + \Gamma_{2p2h} \ ,
\label{eq:new}
\end{eqnarray}
where the $2N$-induced decay rate $\Gamma_{2p2h}$ is assumed to
proceed predominantly through the $\Lambda n p \to n n p$
channel.

On combining the sets of eqs. (\ref{eq:old}) and  (\ref{eq:new})
one finds
\begin{equation}
\left(\frac{\Gamma_n}{\Gamma_p} \right) =
\left(\frac{\Gamma_n^{\rm exp}}{\Gamma_p^{\rm exp}} \right)
\frac{ 1 - \displaystyle\frac{1}{2} \left(
\left(\displaystyle\frac{\Gamma_n^{\rm exp}}{\Gamma_p^{\rm exp}}
\right)^{-1}
+ 1 \right)
 \frac{\displaystyle\Gamma_{2p2h}}{\Gamma_{nm}^{\rm exp}} }
{ 1 - \left(
\displaystyle\displaystyle\frac{\Gamma_n^{\rm exp}}{\Gamma_p^{\rm
exp}} + 1 \right)
 \displaystyle\frac{\Gamma_{2p2h}}{\Gamma_{nm}^{\rm exp}} } \ ,
\label{eq:newrat}
\end{equation}
which gives the new ratio in terms of the old one and the
$2N$-induced decay
rate. From eq. (\ref{eq:newrat}) one can see that if
$\displaystyle\frac{\Gamma_n^{\rm exp}}{\Gamma_p^{\rm exp}} >
0.5$, as is the case of almost all experimental results, then
the ratio $\Gamma_n/\Gamma_p$  of the new analysis is
even larger, in stronger disagreement with the present theories.
Taking as an example the data of \refer{szymanski} for
\hyp{12}{C}, corrected for final state interactions
\cite{franklin}, one extracts
$\displaystyle\frac{\Gamma_n^{\rm exp}}{\Gamma_p^{\rm exp}} =
1.04$  while the new analysis gives
$\displaystyle\frac{\Gamma_n}{\Gamma_p}=1.54$ which shows that,
even if these values have big error bars tied to the
uncertainties in $N_n$ and $N_p$, the $2N$-induced decay channel
is relevant and needs to be considered in any experimental
analysis from where
the ratio $\Gamma_n/\Gamma_p$ is to be extracted.

Note, however, that not all the three nucleons emitted in the
$2N$-induced decay are necessarily fast ones \cite{gal95}. If the
process receives most contributions when the pion is emitted
close to on shell, then the nucleon at the $\Lambda$ vertex will
be
slow and will not be detected due to the experimental energy
tresholds of around $30-40$ MeV. Assuming this to be the case,
the experimental reanalysis gives, instead of the relation
in eq. (\ref{eq:newrat}), the following one
\begin{equation}
\left(\frac{\Gamma_n}{\Gamma_p} \right) =
\left(\frac{\Gamma_n^{\rm exp}}{\Gamma_p^{\rm exp}} \right)
\frac{ 1 - \displaystyle\frac{2}{3} \left(
\left(\displaystyle\frac{\Gamma_n^{\rm exp}}{\Gamma_p^{\rm exp}}
\right)^{-1}
+ 1 \right)
 \displaystyle\frac{\Gamma_{2p2h}}{\Gamma_{nm}} }
{ 1 - \displaystyle\frac{1}{3}\left(
\displaystyle\frac{\Gamma_n^{\rm exp}}{\Gamma_p^{\rm exp}} + 1
\right)
 \displaystyle\frac{\Gamma_{2p2h}}{\Gamma_{nm}} } \ ,
\label{eq:newrat2}
\end{equation}
with $\Gamma_{nm}=\Gamma_n + \Gamma_p + \Gamma_{2p2h}$,
which yields
$\displaystyle\frac{\Gamma_n}{\Gamma_p} <
\displaystyle\frac{\Gamma_n^{\rm exp}}{\Gamma_p^{\rm exp}}$ as
long as
$\displaystyle\frac{\Gamma_n^{\rm exp}}{\Gamma_p^{\rm exp}}<2$.

These considerations lead to the conclusion that, in order to
establish the effect of the $2N$-induced decay mode on the ratio
$\Gamma_n/\Gamma_p$, sensitive to the
detection thresholds, it is necessary to know the spectrum of the
emitted nucleons from the different mechanisms. This is discussed
in the next section.

\subsection{Nucleon spectra}

The findings of the preceding section evidence the need for
theoretical calculations of the energy distribution of nucleons
from the different weak decay mechanisms. These models will also
have to address the problem of final state interactions of the
nucleons on their way out of the nucleus.

The spectra of neutrons and protons from the weak decay of
hypernuclei have been calculated in a recent
work \cite{ramos97}, taking $\Gamma_n/\Gamma_p$ as a free
parameter in order to facilitate the extraction of this ratio
from the experimental data. The spectra were obtained using the
propagator method formalism \cite{oset,ramos94} to determine the
primary emitted nucleons for a given hypernucleus via the LDA, an
approximation particularly suited to treat the
final state interactions,
as we shall see. Let us recall that the rate for a
particular channel
($i=n,p,2N$) is given by
\begin{equation}
\Gamma_i = \int d^3k \tilde{\rho}(\vec{k}) \Gamma_i({\vec k}) =
\int d^3k \tilde{\rho}(\vec{k}) \int d^3r \mid
\phi_\Lambda(\vec{r}) \mid^2 \Gamma_i(\vec{k},\rho(\vec{r}) ) \ ,
\label{eq:lda}
\end{equation}
where $\Gamma_i(\vec{k},\rho(\vec{r}) )$ is the decay rate of a
$\Lambda$ of momentum $\vec{k}$ in nuclear matter of
density $\rho(\vec{r})$, which is the nuclear density at point
$\vec{r}$ where the decay occurs. The function
$\phi_\Lambda(\vec{r})$ is the $\Lambda$ wave function at that
point and an average over the $\Lambda$ momentum distribution is
also implied in eq. (\ref{eq:lda}).
The structure of the integrals involved is then
\begin{equation}
\Gamma_i=\int d^3k \int d^3r \int d^3q \dots {\rm Im}\,
U_i(q^0=k^0-E(\vec{k}-\vec{q}\,)-V_N, \vec{q}\,)
\end{equation}
where the function $U_i$ associated to the particular decay
mechanism can
be further decomposed in terms of the momentum variables of the
emitted nucleons, as explained in \refer{ramos97}. Then, a Monte
Carlo integration method can be used to perform the integrals,
where the random configurations are generated such that: i) give
the relative amount of one-nucleon- and two-nucleon-induced decay
probabilities according to the model described in Sect. 3.3, and
ii) give the relative amount of neutron- and proton-induced decay
rates, according to the input value of $\Gamma_n/\Gamma_p$, used
as a parameter of the calculation.

In this way, each configuration determines, at each point
$\vec{r}$, a number of primary emitted nucleons and their
corresponding momenta. The fate of these nucleons on their way
out of the nucleus is then followed using a Monte Carlo
simulation method \cite{Ca94}, which allows the nucleons to
undergo collisions with other nucleons according to $NN$ cross
sections modified by Pauli blocking. In each collision, the
nucleons change energy, direction and, eventually, charge, as
well as exciting other nucleons above the local Fermi momentum.
Then, the propagation of these secondary unbound nucleons through
the nucleus is also followed.

\begin{figure}[hbt]
       \setlength{\unitlength}{1mm}
       \begin{picture}(100,100)
      \put(30,5){\epsfxsize=10cm \epsfbox{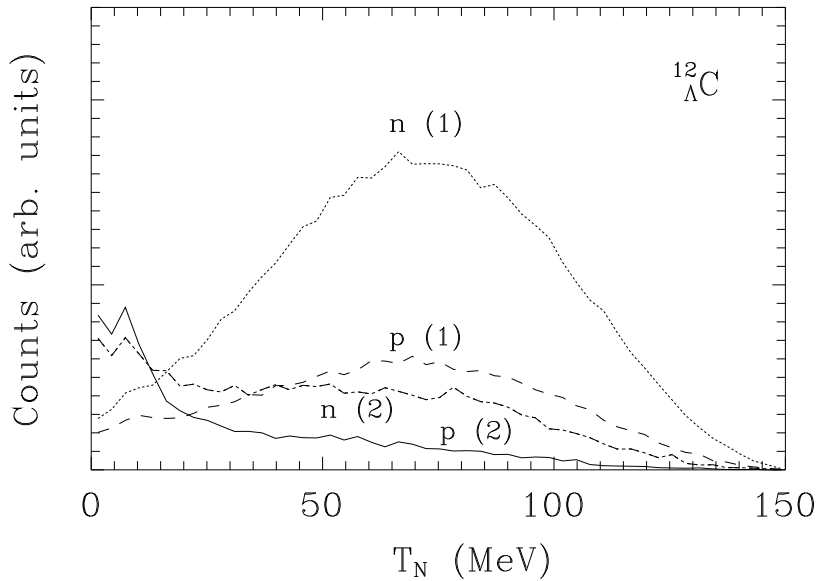}}
       \end{picture}
\caption{ Spectra of neutrons and protons in the decay of
\hyp{12}{C}, assuming a ratio $\Gamma_n/\Gamma_p=1$. Dashed line:
protons from $1N$-induced decay.
Dotted line: neutrons from $1N$-induced decay.
Solid line: protons from $2N$-induced decay.
Dash-dotted line: neutrons from $2N$-induced decay.
}
\label{fig:spec}
\end{figure}

The obtained spectra of neutrons and protons coming from the
$1N$- and $2N$-induced mechanisms in the decay of \hyp{12}{C}
are shown in fig. \ref{fig:spec} for a value
$\Gamma_n/\Gamma_p=1$.
One observes that the distribution of protons (and neutrons) from
the $1N$-induced mechanism peaks around $70-75$ MeV, which
corresponds to
the energy of the most probable kinematics in which the two
nucleons emerge back-to-back. The energy distribution
generated by to the $2N$-induced process is flatter and presents
a peak
at low excitation energies, which contains mainly slow nucleons
coming from the $\Lambda$ vertex plus some nucleons whose energy
has been degraded by the effect of final state interactions.
It is clear then that the experimental thresholds ($30-40$ MeV)
will prevent these $2N$-induced nucleons from being detected.
However, a considerable fraction still leaks into the region
where most $1N$-induced decay nucleons appear. Hence,
separating the nucleons from both mechanisms will be very
difficult, unless angular correlation measurements are also
conducted. This effort is presently being carried out at
Brookhaven and it is certainly encouraging to see that their
preliminary results \cite{zeps97} already show how the
back-to-back kinematics nicely selects the nucleons emitted in
the $1N$-mechanism and, therefore, a cleaner extraction of the
ratio
$\Gamma_n/\Gamma_p$ might soon become available.

Qualitatively similar spectrum shapes were obtained for
\hyp{5}{He} in \refer{shinmu97}, where the $2N$-induced mechanism
was assumed
to proceed via a $\Delta$-excited intermediate state, although
the medium modification of the $\Delta$ and other important
$2p2h$ mechanisms were ignored. Final state interactions were not
considered, which is reasonable for the light hypernucleus
treated but does not allow one to apply the model to heavier
hypernuclei.

It is clear that the spectra shown in fig. \ref{fig:spec} would
certainly differ if other values of
the ratio $\Gamma_n/\Gamma_p$ were used. A comparison of the
available experimental proton spectra \cite{szymanski,montwill}
with the calculated ones for several values of
$\Gamma_n/\Gamma_p$ was also done in \refer{ramos97} and it was
found that values of $\Gamma_n/\Gamma_p\sim 2-3$ were favored, in
strong contradiction with the OPE models.

However, in the same work, other observables that were less
sensitive to the specific details of the Monte Carlo simulation
determining the final shape of the spectrum were also studied as
functions of $\Gamma_n/\Gamma_p$, which would probably lead to a
more reliable determination of this ratio.
By integrating over the energy spectra, the total number of
neutrons, $N_n$, and protons, $N_p$, and therefore the ratio
$N_n/N_p$ was plotted against
$\Gamma_n/\Gamma_p$. The experimental values of
\refer{szymanski}, corrected in \refer{franklin}, give
$N_n/N_p=(2530 \pm 1050)/(1112 \pm 130)$. The error band in
$N_n/N_p$ allowed values
for $\Gamma_n/\Gamma_p$ in the range $0.0-1.65$ \cite{ramos97},
perfectly compatible with the OPE predictions. Note that the
inclusion of the $2N$-induced channel extended the possible
values for $\Gamma_n/\Gamma_p$ at both ends of the interval, with
respect to what would have been obtained omitting this channel.

\begin{figure}[htb]
       \setlength{\unitlength}{1mm}
       \begin{picture}(100,140)
      \put(35,5){\epsfxsize=8cm \epsfbox{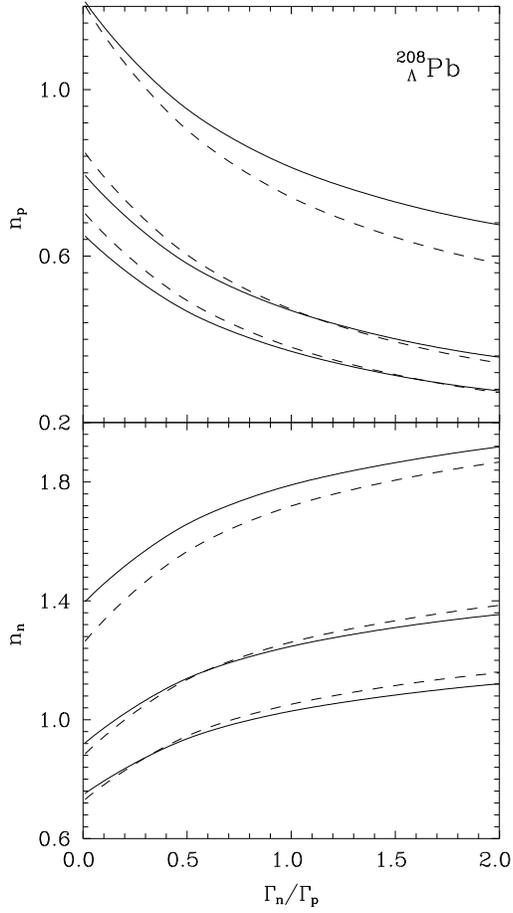}}
       \end{picture}
\caption{
Number of protons ($n_p$) and neutrons ($n_n$) per $\Lambda$
decay event in \hyp{208}{Pb}. From top to bottom the results
include energy cuts of 0, 30 and 40 MeV. Dashed lines: $1N$-
induced mechanism. Solid lines: $1N + 2N$ mechanisms.
}
\label{fig:nnnp}
\end{figure}

The fact that the relative error of $N_n/N_p$ is the sum of the
relative errors of $N_n$ and $N_p$, together with the fact that
neutrons are measured with little precision, makes the
uncertainty on $N_n/N_p$ very large and therefore leads to a poor
determination of
$\Gamma_n/\Gamma_p$. It was pointed out in \refer{ramos97} that
the separate measurement of the number of protons and neutrons
per decay event, $n_p$ and $n_n$, would provide more reliable
predictions. As an
example, fig. \ref{fig:nnnp} shows the number of protons
(upper half) and neutrons (lower half) per decay event
for different energy cut-offs (from top to bottom: 0, 30 and 40
MeV) in the decay of \hyp{208}{Pb}. If one considers a
hypothetical situation in which the values of $n_p$ and $n_n$
could be measured with 10\% accuracy and the value of
$\Gamma_n/\Gamma_p$ was around 1, then this would imply having
measured
values of $n_p=0.37 \pm 10\%$ and $n_n=1.03 \pm 10\%$, assuming
an energy cut-off of 40 MeV.
Representing these values in the figure
and translating their error bands into the corresponding
ones for $\Gamma_n/\Gamma_p$, one would obtain, from $n_p$, the
value
$\Gamma_n/\Gamma_p=1.0 ^{+0.35}_{- 0.22}$ and, from $n_n$, the
value
$\Gamma_n/\Gamma_p=1.0^{+1.0}_{-0.52}$. This example clearly
demostrates that the measurement of $n_p$ is, so far, the most
crucial one in order to determine the ratio $\Gamma_n/\Gamma_p$.
An experimental set up to be used at TJNAF and at the Yerevan
Electron Synchroton (YESS) facility has already been devised in
order to measure $n_p$ from the delayed fission events from the
decay of heavy hypernuclei \cite{bayatyan}, which hopefully will
lead to a cleaner determination of the ratio
$\Gamma_n/\Gamma_p$.

\section{Conclusions and perspective}

The discussions along this review have shown some interesting
features that we summarize here.

The mesonic decay of $\Lambda$ hypernuclei has offered some
evidence of the strong repulsion at short distances of the
$\Lambda N$ interaction which appears naturally in current models
of the $YN$ interaction. The sensitivity of the mesonic decay to
the pion nucleus interaction has been stressed repeatedly in the
calculations. The total mesonic decay rate is appreciably
enhanced due to the p-wave attraction of the optical potential.
However, exclusive mesonic decay into particular final nuclear
states may select the s-wave part and
the partial rate can be reduced, particularly if the final
nucleus is a closed shell one. Thus, investigations on $\Lambda$
mesonic decay into different channels can offer complementary
information on
the pion nucleus optical potential and should help in deciding
between different parametrizations of theoretical models, which
are equally successful in explaining pion nuclear scattering or
pionic atoms data.
Moreover, the mesonic decay rates of heavy hypernuclei, even if
small, would be most welcome since the pion renormalization is
most important there. It should also be stressed that these
measurements would also provide information about the $\pi^0$
nuclear interaction which is not available from scattering data.

The non-mesonic decay has been the object of intensive
theoretical investigations and models going beyond the
traditional one pion exchange mechanism are now available. Yet,
the inclusion of heavier mesons or multimeson exchange does not
change appreciably neither the decay rate nor the ratio
$\Gamma_n/\Gamma_p$, which was the main motivation to develop
such models. The incorporation of quark degrees of freedom into
the picture brings interesting new features, such as important
$\Delta I=3/2$ contributions to the decay rate and the
possibility of obtaining increased values for the
$\Gamma_n/\Gamma_p$ ratio. However, more theoretical
investigations are
needed to understand both the mesonic and the non-mesonic decays
from the same basic quark Hamiltonian. Although the discrepancy
between the theoretical models and experiment for the value of
$\Gamma_n/\Gamma_p$ is still a puzzle, one must note that the
experimental uncertainties for this ratio are very large. The
two-nucleon induced decay was proposed as a hope to resolve this
puzzle but, as discussed here, the consideration of this channel
leads to enlarged error bars in the experimental analysis.
Nevertheless, some efficient methods to determine
$\Gamma_n/\Gamma_p$ with far more precision than has been
possible until now have been proposed. One possibility is through
the determination of the number of protons per $\Lambda$ decay
and their spectra. New experiments along these lines are now
planned and most awaited. Another interesting feasible experiment
is the measurement in coincidence of two nucleons emerging
back-to-back. This should allow to disentangle the nucleons of
the one-nucleon-induced mechanism from those of the
two-nucleon-induced one, leading to
a cleaner determination of the $\Gamma_n/\Gamma_p$ ratio.

Measurements of the lifetime of heavy hypernuclei are now
available from different experiments and, although agreeing at
the qualitative level, more precise determinations of these rates
would be also helpful to establish clearly when saturation is
reached and, therefore, obtain information on the range of the
weak $\Lambda N \to NN$ transition.

The $\Delta I=1/2$ violation in the non-mesonic weak decay of
hypernuclei is an intriguing possibility that should be explored
further through ratios of partial decay rates of light
hypernuclei.

The possibility of producing polarized hypernuclei with the
$(\pi^+,K^+)$ reaction has extended the available observables to
asymmetries of the weak decay products. These results bring
complementary information to that obtained from partial and total
non-mesonic decay rates of hypernuclei and some theoretical
models have
now predictions for the asymmetry of the emitted particles.
However, in order to constrain
these models, it is necessary to reduce the experimental
uncertainties and clarify the contradicting experimental results.

In summary, the field of $\Lambda$ hypernuclei and their 
weak decay modes has
experienced an impressive progress in the last decade. The
availability of new data from practically all intermediate energy
experimental facilities is building up the grounds for a bright
future in which one may expect to settle the present problems and
open the door to novel an interesting physics.
\vspace{2cm}

{\large\bf Acknowledgements}

\vskip 0.5 cm

This work has been partially supported by DGICYT contracts
PB95-1249 and PB96-0753, and by the E.U. contract CHRX-CT93-0323.

We would like to thank our colleagues C. Bennhold, A. Faessler,
P. Fern\'andez de C\'ordoba, C. Garc\'{\i}a-Recio, J. Nieves, A.
Parre\~no, L.L. Salcedo, U. Straub and M.J. Vicente-Vacas, who
collaborated in obtaining many of the results reported here.

\end{document}